\documentclass{emulateapj}

\usepackage{graphicx}
\usepackage[hypertex]{hyperref}

\def \nh {N${\rm _H}$}
 
\def \arcmin {\hbox{$^\prime$}} 
\def \arcsec {\hbox{$^{\prime\prime}$}} 
 
\def\spose#1{\hbox to 0pt{#1\hss}} 
\def\ltsim{$\mathrel{\spose{\lower 3pt\hbox{$\sim$}} 
        \raise 2.0pt\hbox{$<$}}$\thinspace} 
\def\gtsim{$\mathrel{\spose{\lower 3pt\hbox{$\sim$}} 
        \raise 2.0pt\hbox{$>$}}$\thinspace} 
\def \msun {${\rm M_\odot}$} 
 
\def \nh {$N_{\rm H}$}

\def \eg {e.g.}

\def \dtwentyfive {${\rm D_{25}}$}
\newcommand{\rhog}{${\rm \rho_g}$}

\newcommand{\mvir}{${\rm M_{vir}}$}
\newcommand{\reff}{${\rm R_{eff}}$}
\newcommand{\mstars}{${\rm M_{*}}$}

\newcommand{\sigmaeight}{$\sigma_8$}
\newcommand{\sersic}{S\'{e}rsic}
\newcommand{\apec}{APEC}

\newcommand{\zfe }{${\rm Z_{Fe}}$}
\newcommand{\zo }{${\rm Z_{O}}$} 
\newcommand{\zsi }{${\rm Z_{Si}}$}
\newcommand{\zni }{${\rm Z_{Ni}}$}
\newcommand{\zs }{${\rm Z_{S}}$}
\newcommand{\zne }{${\rm Z_{Ne}}$}
\newcommand{\zmg }{${\rm Z_{Mg}}$}

\newcommand{\mgrav}{${\rm M_{grav}}$}
\newcommand{\chandra }{{\em Chandra}} 
\newcommand{\xspec }{{\em Xspec}} 
\newcommand{\acis }{{\em ACIS}} 
\newcommand{\ciao }{{\em CIAO}} 
\newcommand{\caldb }{{\em Caldb}} 
\newcommand{\heasoft }{{\em Heasoft}}
\newcommand{\lcdm}{$\Lambda$CDM}
\newcommand{\ergps}{${\rm erg\ s^{-1}}$} 
\newcommand{\xmm }{{\em XMM}} 
\newcommand{\asca }{{\em ASCA}} 
\newcommand{\rosat }{{\em ROSAT}}

\newcommand{\lx }{${\rm L_X}$}
\newcommand{\lb }{${\rm L_B}$}
\newcommand{\lk }{${\rm L_K}$}
\newcommand{\ks }{${\rm K_s}$}
\newcommand{\rvir }{${\rm R_{vir}}$}
\newcommand{\lstar}{${\rm L_{*}}$}
\newcommand{\twomass}{{\em 2MASS}}

\newcommand{\fbaryons}{${\rm f_{b}}$}
\newcommand{\lsun }{${\rm L_\odot}$}
\newcommand{\leda}{{\em{LEDA}}} 
\newcommand{\ned}{{\em{NED}}} 

\newcommand{\thin}{\thinspace}
\slugcomment{Submitted to the Astrophysical Journal}
\shorttitle{A Chandra View of Dark Matter in Galaxies.}
\shortauthors{Humphrey et al.}
 
\begin{document} 
 
\title{A Chandra View of Dark Matter in Early-Type Galaxies}
\author {\href{mailto:phumphre@uci.edu}{Philip J. Humphrey}\altaffilmark{1}, David A. Buote\altaffilmark{1}, Fabio Gastaldello\altaffilmark{1}, Luca Zappacosta\altaffilmark{1}, James S. Bullock\altaffilmark{1}, Fabrizio Brighenti\altaffilmark{2,3} and William G.~Mathews\altaffilmark{3}}
\altaffiltext{1}{Department of Physics and Astronomy, University of California at Irvine, 4129 Frederick Reines Hall, Irvine, CA 92697-4575}
\altaffiltext{2}{Dipartimento di Astronomia, Universit\`{a} di Bologna, Via Ranzani 1, Bologna 40127, Italy}
\altaffiltext{3}{University of California Observatories, Lick Observatory, University of California at Santa Cruz, Santa Cruz, CA 95064}
\begin{abstract}
We present a \chandra\ study of mass profiles in 
7 elliptical galaxies, of which 3 have galaxy-scale and 
4 group-scale halos, demarcated at $10^{13}$\msun.
These represent the best available data for 
nearby objects with comparable X-ray luminosities.
We measure $\sim$flat mass-to-light (M/L) profiles
within an optical half-light radius (\reff), rising by an order
of magnitude at $\sim$10\reff, which confirms the presence of 
dark matter (DM).
The data indicate hydrostatic equilibrium, which is 
also supported by agreement with studies of stellar kinematics
in elliptical galaxies.
The data are well-fitted by a model comprising an NFW DM profile
and a baryonic component following the optical light.
The distribution of DM halo concentration 
parameters (c) {\em versus} \mvir\ agrees with
\lcdm\ predictions and our observations of bright groups. 
Concentrations are slightly higher than expected,
which is most likely a selection effect.
Omitting the stellar mass drastically increases c,
possibly explaining large concentrations found by some
past observers.
The  stellar M/\lk\ agree  with population synthesis models, 
assuming a Kroupa IMF. Allowing adiabatic compression (AC)
of the DM halo by baryons made M/L more discrepant, casting
some doubt on AC.
Our best-fitting models imply total baryon fractions $\sim$0.04--0.09,
consistent with models of galaxy formation incorporating strong
feedback.
The groups exhibit positive temperature gradients, consistent
with the ``Universal'' profiles found in other groups and clusters,
whereas the galaxies have negative gradients, 
suggesting a change in the evolutionary history of the systems
around \mvir$\simeq 10^{13}$\msun.
\end{abstract}

\keywords{Xrays: galaxies--- galaxies: elliptical and lenticular, cD--- 
galaxies: halos--- galaxies: ISM--- dark matter}

\section{Introduction}
The nature and distribution of dark matter (DM) in the Universe is one
of the fundamental problems facing modern physics. Cold DM lies at the
heart of  our  current (\lcdm) cosmological   paradigm, which predicts
substantial  DM halos for objects  at all mass-scales from galaxies to
clusters.   Although  \lcdm\  has  been    remarkably  successful   at
explaining large-scale features
\citep[\eg][]{spergel03a,perlmutter99a}, 
observations of galaxies have been more problematical for the theory.
Dissipationless dark matter simulations find that dark matter
halos are well characterized by a
``Universal'' mass density profile \citep[][hereafter NFW]{navarro97}
over a wide range of Virial masses (\mvir) \citep[e.g.][]{bullock01a}.
Low mass halos tend to form first in hierarchical cosmologies
and are consequently more tightly concentrated than their
later forming,  high mass counterparts.
This tendency  produces a predicted correlation
between the DM halo concentration parameter (c, which is ratio between 
Virial radius, \rvir, and the characteristic scale of the density profile)
and \mvir \citep{navarro97}.  However, since mass and formation
epoch are not perfectly correlated, we expect a significant
scatter at fixed Virial mass \citep{jing00a,bullock01a,wechsler02a}.
The tight link between halo formation epoch and concentration
implies that the precise relation between c and \mvir\
is sensitive to the underlying Cosmological parameters,
including \sigmaeight\ and the dark energy equation of state
\citep{kuhlen05a}, making an observational test of this relation
potentially a very powerful tool for cosmology.

The mass profiles of galaxies also may provide valuable clues
as to the way in which galaxies form  in DM halos. 
In particular, as baryons cool and collapse into stars, the 
associated increase in the central mass density should in turn
modify the shape of the DM halo. This process is typically 
modelled assuming adiabatic contraction (AC) of the DM particle orbits
\citep[\eg][]{blumenthal86a,gnedin04a}. 
If the galaxy halo subsequently evolves by major mergers,
simulations are unclear as to whether these features would persist 
\citep[\eg][]{gnedin04a} or whether the merging process may
destroy this imprint of star formation, or even 
mix the DM and baryons sufficiently to produce a {\em total}
gravitating mass profile more akin to NFW \citep{loeb03a,elzant04a}.

Observational tests of the predictions of \lcdm\ have proven
controversial. In clusters of galaxies there is overwhelming evidence
for DM, and an increasing body of work verifying the 
predictions of \lcdm. In particular recent, high-quality 
\chandra\ and \xmm\ observations have revealed mass profiles
in remarkable agreement with the Universal profile from
deep in the core to a large fraction of 
\rvir\ \citep[\eg][]{lewis03a,zappacosta06a,vikhlinin05b},
and a distribution of c {\em versus} \mvir\ in good agreement with
\lcdm\ \citep{pointecouteau05a}.
In galaxies, however, the picture is much less clear. Rotation curve
analysis of low surface brightness (LSB) disk galaxies has suggested
significantly less cuspy density profiles than expected 
\citep[\eg][]{swaters00a}. 
Although this discrepancy led to a serious discussion of 
modifications to the standard paradigm 
\citep[\eg][]{hogan00a,spergel00a,zentner02a,kaplinghat05a,cembranos05a}, 
recent  results, taking account of observational bias and the 3-dimensional
geometry of the DM halos, have done much to resolve the discrepancy
\citep[\eg][]{swaters03a,simon05a}. However, some 
significant discrepancies remain, not least of which 
is that the DM halos of these galaxies appear less concentrated
than expected \citep[\eg][]{gonzalez00a,kassin06a}.
A possible explanation is that LSB galaxies are preferentially 
found in low-concentration halos \citep{bullock01a,bailin05a,wechsler05a}, making
additional constraints at the galaxy scale extremely important. %DONE

In many respects, kinematical mass measurements 
are far more challenging for early-type than spiral galaxies.
As essentially pressure-supported systems little is known
{\em a priori} about the velocity anisotropy tensor of
the stars in elliptical galaxies, which is problematical for
the determination of the mass from stellar motions.
Nonetheless, stellar kinematical measurements
have widely been used as a means to measure the gravitating matter
within $\sim$the optical half-light radius (\reff) of 
elliptical galaxies \citep[\eg][]{binney90a,vandermarel91a,gerhard01a}.
These studies tend to find
relatively flat mass-to-light (M/L) ratios within \reff, 
implying that most of the matter within this radius is baryonic.
Consideration of the tilt in the fundamental plane can also
lead to the same conclusion \citep{borriello03a}.
In contrast, \citet{padmanabhan04a} pointed out that 
dynamical M/L ratios within \reff\ are much larger than 
predicted by realistic stellar population synthesis models for 
stars alone, allowing 
\gtsim 50\% of the mass within \reff\ to be dark. %DONE

Attempts to extend kinematical studies of elliptical galaxies
to larger radii, where DM should be dominant, 
have proven controversial. 
In particular \citet{romanowsky03a} argued against the existence
of DM in a small sample of elliptical galaxies, based on 
planetary nebulae dynamics within $\sim$5\reff. 
We note that this sample was heavily biased towards very X-ray
faint objects, which might hint at low-mass halos
since they have not held onto their hot gas. 
In any case \citet{dekel05a} pointed out that their
conclusions were very sensitive
to the uncertainty in the velocity anisotropy tensor,
for plausible values of which the data were consistent
with substantial DM halos.
In fact globular cluster dynamics in one of these systems,
NGC\thin 3379, does imply a significant amount of DM 
\citep{pierce06a,bergond06a}.
As more kinematical studies of early-type galaxies at
large radii are appearing, it is becoming clear
that at least some elliptical galaxies host considerable
DM halos \citep[\eg][]{statler99a,romanowsky05a}. There
persist some questions, however, as to the extent 
to which all galaxies have DM halos consistent with \lcdm.
In particular \citet{napolitano05a} argued
that a substantial number of early-type galaxy halos
appear  less concentrated than expected.

Gravitational lensing provides further evidence
that, at least some, early-type galaxies possess substantial
DM halos \citep[\eg][]{kochanek95a,fischer00a,rusin02a}. 
Since weak lensing of galaxies only provides useful mass constraints
in a statistical sense, the relatively rare instances of 
strong lensing are required to study DM in individual systems.
Nonetheless it has been possible in a few cases to decompose 
the mass into stellar and DM components, albeit with strong 
assumptions or additional observational constraints
\citep[\eg][]{rusin03a,treu04a}.

X-ray observations of the hot gas in early-type galaxies provide a 
complementary means to infer the mass-profiles {\em via}
techniques similar to those used in studying clusters. 
Since the X-ray emission from early-type galaxies is typically
not very bright, prior to the 
advent of  \chandra\ and \xmm\ this was limited
by the relatively sparse information on the 
radial temperature and density profiles of the hot gas
which could be determined by prior generations of 
satellites.
Notwithstanding this limitation, large M/L ratios (consistent
with substantial DM) were 
inferred for a number of X-ray bright galaxies,
albeit with strong assumptions concerning the temperature and density
profiles \citep[\eg][]{forman85a,loewenstein99b}.
Using a novel technique which relied, instead, on the
ellipticity of the X-ray halo, \citet{buote94} were able robustly to 
detect DM  in the isolated elliptical
NGC\thin 720 \citep[see also][]{buote96a,buote98d,buote02b}. 
Detailed measurements of the radial mass distribution 
were, however, largely restricted
to a few massive systems, which may be entwined with a group
halo \citep[\eg][]{irwin96,brighenti97a}. 
Nevertheless \citet{brighenti97a}
were able to decompose the mass profiles of two systems, NGC\thin 4472
and NGC\thin 4649, into stellar and DM components.
\citet{sato00a} investigated the \mvir-c relation using \asca\ for a 
sample of objects ranging from massive clusters to $\sim$3 elliptical
galaxies. The limited spatial resolution of \asca\ necessitated
some assumptions about the density  profiles
and, crucially, the authors neglected any stellar mass component
in their fits. This omission may explain the very steep
\mvir-c relation (with c$_{200}$\gtsim 30 for the galaxies) found by
these authors, in conflict with \lcdm\ \citep{mamon05a}.

Although mass profiles of early-type galaxies are beginning to appear
which exploit the improved sensitivity and resolution of \chandra\ and \xmm,
many of the most interesting constraints on DM are still restricted to
massive systems, which may be at the centres of groups.
For example, \citet{fukazawa06a} reported \chandra\ and \xmm\ M/L profiles for 
$\sim$50 galaxies and groups,
confirming $\sim$flat profiles within \reff\ which rise at larger
radii. However, the constraints at large radii were dominated by
the massive (group-scale) objects so the implications for the DM
content of normal galaxies are unclear. Furthermore,
the authors  included a substantial number of 
highly disturbed systems, in which hydrostatic equilibrium
may be questioned, and failed to account for the unresolved
sources which dominate the emission in the lowest-\lx\ objects
in their sample\footnote{Although the authors account for unresolved
sources when measuring the gas temperature, they do not account for
it when computing the gas density, where its effect is more pronounced}.
Recently, however, detailed \chandra\ and \xmm\ mass profiles have begun to
appear for isolated early-type galaxies, also confirming the presence
of massive DM halos \citep[\eg][]{osullivan04b,khosroshahi04a}.

This paper is part of a series 
\citep[see also][]{gastaldello06a,zappacosta06a,buote06b,buote06a}
using high-quality \chandra\ and \xmm\ data to investigate the 
mass profiles of  galaxies, groups and clusters. This provides
an unprecedented opportunity to place definitive constraints upon 
the \mvir-c relation over $\sim$2 orders
of magnitude in \mvir. In this paper, we focus on the temperature,
density and mass profiles of seven galaxies and poor groups 
chosen from the \chandra\ archive.
In order to compare to theory we perform spherically-averaged
analysis, leaving a discussion of the ellipticities of the 
X-ray halos to a future paper.
In \S~\ref{sect_targets} we discuss the target selection. The 
data-reduction is described in \S~\ref{sect_reduction} and 
the X-ray morphology is addressed in \S~\ref{sect_imaging}.
We discuss the spectral analysis in \S~\ref{sect_spectra}, the
mass analysis in \S~\ref{sect_mass}, the systematic uncertainties
in our analysis in \S~\ref{sect_systematics}
and reach our conclusions in \S~\ref{sect_discussion}.
The three systems for which we find \mvir$<10^{13}$\msun\ are
optically isolated and so we refer to them as ``galaxies'', and
the other systems in our sample as groups. We discuss this in
more detail in  \S~\ref{discussion_groups}.
In this paper, all error-bars quoted represent 90\% confidence limits, 
unless otherwise stated, and we computed Virial quantities assuming 
a ``critical overdensity''
factor for the DM halos of $\rho_{\rm halo}/\rho_{\rm crit} = 103$
(where $\rho_{\rm halo}$ is the mean density of a sphere of 
mass \mvir\ and radius \rvir).

\section{Target Selection} \label{sect_targets}
We chose, for this initial study, to focus on objects observed with
\chandra. \chandra\ data are particularly valuable
for the study of galaxies since the unprecedented spatial resolution
makes it possible to resolve the temperature and density profiles
deep into the galaxy core, allowing us to disentangle the stellar
and dark matter, and resolve them into discrete components.
We initially chose a set of potential target systems from detections
listed in the X-ray
catalogue of \citet{osullivan01a} which have non-grating ACIS
data in the \chandra\ archive. To eliminate bright 
groups and cluster cDs in the sample, we excluded galaxies with 
\lx\gtsim $10^{43}$\ergps. In order to perform the required spatially-resolved
spectroscopy, we required at least $\sim$5000 hot gas photons. 
The potential targets were processed
and the 0.1--10.0~keV image examined for evidence of large-scale
disturbances (\S~\ref{sect_imaging}). We included some systems
with low-amplitude asymmetries which should not strongly 
disturb hydrostatic equilibrium (we discuss this in more detail in
\S~\ref{sect_asymmetry}).
Preliminary analysis was conducted to estimate the Virial mass of 
the object (\S~\ref{sect_mass}). Since we aimed to focus on
lower-mass objects, systems for which a fit
using a simple NFW profile yielded \mvir\gtsim $10^{13}$\msun were
discounted. Massive objects of this type are the focus of another study
\citep{gastaldello06a}. The most promising
candidates for study found {\em via} this method were chosen for detailed
analysis. The properties of the 7 objects in our sample
and the \chandra\ exposures are shown in Table~\ref{table_obs}.

Our selection criteria naturally bias the sample towards
X-ray bright galaxies. One might expect that galaxies sitting in 
deep potential wells are more likely to retain hot gas than those 
with little dark matter, and so our results may be biased somewhat 
towards those galaxies with substantial dark halos \citep[in contrast
to the opposite bias in the analysis of][]{romanowsky03a}.
As we are selecting objects which are not heavily disturbed, we are 
also biased towards galaxies which have not recently undergone a 
major merger. For the purposes of this paper, however, we do not 
require statistical completeness, and we will discuss how to 
take account of these selection effects in \citet{buote06b}.
%\clearpage
\begin{deluxetable*}{lllllllrrr}
\tablecaption{The galaxy sample\label{table_sample}}
\tabletypesize{\scriptsize}
\tablehead{
\colhead{Galaxy} & \colhead{Type} & \colhead{\lb}  & \colhead{\lk} & \colhead{Dist} & 
\colhead{Scale} & \colhead{\reff}
& \colhead{ObsID} & \colhead{Date} & \colhead{Exposure} \\
\colhead{} & \colhead{} & \colhead{($10^{10}$\lsun)}  &\colhead{($10^{11}$\lsun)}  & \colhead{(Mpc)} & \colhead{(\arcsec\ kpc$^{-1}$)} & \colhead{(kpc)} & 
\colhead{} & \colhead{(dd/mm/yy)} & \colhead{(ks)}
}
\startdata
NGC\thin 720  & E5            & 3.1  & 1.7 & 25.7 & 8.1 & 3.1 & 492 & 12/10/00 & 17 \\
NGC\thin 1407 & E0            & 6.4  & 3.1 & 26.8 & 7.8 & 4.4 & 791 & 16/08/00 & 38 \\
NGC\thin 4125 & E6 pec Liner  & 4.7  & 1.8 & 22.2 & 9.4 & 3.3 & 2071& 09/09/01 & 63 \\
NGC\thin 4261 & E2-3 Liner Sy3& 4.4  & 2.2 & 29.3 & 7.1 & 3.4 & 834 & 06/05/00 & 34 \\
NGC\thin 4472 & E2/S0(2) Sy2  & 7.5  & 3.2 & 15.1 & 14  & 4.0 & 321 & 12/06/00 & 34 \\
NGC\thin 4649 & E2            & 5.1  & 2.5 & 15.6 & 13  & 3.2 & 785 & 20/04/00 & 21 \\
NGC\thin 6482 & E Liner       & 10.9 & 3.2 & 58.8 & 3.6 & 3.4 & 3218& 20/05/02 & 18 
\enddata
\tablecomments{The galaxies in the sample. Distances were obtained from 
\citet{tonry01}, corrected for the the new Cepheid zero-point 
\citep{jensen03},
except for NGC\thin 6482, for which we adopted the kinematical distance
modulus from \leda. \lb\ was obtained from \leda, corrected to our distance.
\ks-band luminosities (\lk) and 
effective radii (\reff) were obtained from \twomass. We assumed 
${\rm M_{B\odot}=5.48}$  and ${\rm M_{K\odot}=3.41}$
\citep[\eg][]{maraston98a}. We also list the image scale
(Scale), which is the number of arc seconds corresponding to 1 kpc.
We list the observation ID (ObsID) and total exposure times, after
having eliminated flaring intervals.}\label{table_obs}
\end{deluxetable*}
%\clearpage
\section{Data reduction} \label{sect_reduction}
For data reduction, we used the \ciao\ 3.2.2 and \heasoft\ 5.3 software
suites, in conjunction with \chandra\ calibration database (\caldb)
version 3.1.0. Spectral-fitting was conducted with \xspec\ 11.3.1w.
In order to ensure the most up-to-date calibration, all data were 
reprocessed from the ``level 1'' events files, following the standard
\chandra\ data-reduction threads\footnote{\href{http://cxc.harvard.edu/ciao/threads/index.html}{http://cxc.harvard.edu/ciao/threads/index.html}}.
We applied the standard correction to take account of the time-dependent gain-drift 
as implemented in the standard \ciao\ tools. To identify periods of enhanced
background (``flaring''), which seriously degrades the signal-to-noise (S/N)
and complicates background subtraction \citep{markevitch02}
we accumulated background lightcurves for each exposure from
low surface-brightness regions of the active chips. We
excluded obvious diffuse emission and data in the vicinity of any detected
point-sources (see below). Periods of flaring were identified by eye and
excised. Small amounts of residual flaring not removed by this procedure 
can be important in low surface-brightness regions at large radii, 
but this was taken into account in our treatment of the 
background (\S~\ref{sect_bkd}).
The final exposure times are listed in Table~\ref{table_obs}.

Point source detection was performed using the \ciao\ tool
{\tt wavdetect} \citep{freeman02}. Point sources were identified 
in full-resolution images of the \acis\ focal-plane, containing all active
chips (except the S4 chip, which suffers from serious ``streaking'', which can
lead to false detections). To maximise the likelihood of identifying
sources with peculiarly hard or soft spectra, images were created in three
energy bands, 0.1--10.0~keV, 0.1--3.0~keV and 3.0--10.0~keV. Sources were
detected separately in each image. In order to minimize spurious detections at
node or chip boundaries we supplied the detection algorithm with
exposure-maps  generated at
energies 1.7~keV, 1.0~keV and 7~keV respectively (although the precise
energies chosen made little difference to the results). The
detection algorithm searched for structure over pixel-scales of 1, 2, 4, 8 and
16 pixels, and the detection threshold was set to ensure $\sim$0.1 
spurious detections per image.
The source-lists obtained within each energy-band were combined and
duplicated sources removed, and the final list was checked
by visual inspection of the images.
The data in the vicinity of any detected point source
were removed so as not to contaminate the diffuse emission.
As discussed in \citet[][see also \citealt{kim03a}]{humphrey04a}
a significant fraction of faint X-ray binary sources
will not have been detected by this procedure, and so we include
an additional component to account for it in our spectral fitting
(\S~\ref{sect_spectra}).

For each galaxy, we extracted spectra in a number
of concentric annuli, centred on the nominal X-ray centroid. We determined
the centroid iteratively by placing a 0.5\arcmin\ radius aperture at the nominal
galaxy position (obtained from \ned) and computing the X-ray centroid
within it. The aperture was moved to the newly-computed centroid, and the
procedure repeated until the computed position converged. Typically the
X-ray centroid agreed with that from \ned. The widths of
the annuli were chosen so as to contain approximately the same number of
background-subtracted photons and ensure there were sufficient photons in each
to perform useful spectral-fitting. The data in the vicinity of any
detected point-sources were excluded, as were the data from the vicinity
of chip gaps, where the instrumental response may be uncertain. We extracted
products from all active chips, excluding the S4, since it suffers from
considerable ``streaking'' noise. 
Appropriate count-weighted spectral response 
matrices were generated for each annulus 
using the standard \ciao\ tasks {\bf mkwarf} and {\bf mkacisrmf}.

\subsection{Background estimation} \label{sect_bkd}
One of the chief difficulties in performing spectral-fitting of 
diffuse emission is the proper treatment of the background.
A set of standard blank-field ``template'' files are available for
\chandra\ as part of the \caldb. We found, however, that 
the background template files are not sufficiently accurate to use 
in the very low surface brightness regions at large radii, which are
crucial to determine interesting global mass constraints.
The background comprises cosmic, instrumental and non X-ray
(particle) components.
The cosmic  component is known to vary  from field to field, while
the non X-ray background exhibits long-term secular variability.
To mitigate the latter effect, several authors have
adopted the practice of renormalizing the background
template to ensure good agreement with their data at high
energies (\gtsim 10~keV). Such a procedure, however, also
renormalizes the (uncorrelated) cosmic X-ray background and instrumental
line features, which can lead to serious over- or under-subtraction.
Given these reservations we chose to use an alternative
background estimation procedure.

Our  method involved modelling the background, somewhat akin to
the approach of  \citet{buote04c}. All of the targets were centred on the 
\acis-S3 chip, which is back-illuminated (BI). To obtain constraints on the 
background, we extracted spectra from a 
$\sim$2\arcmin\ region centred on the S1 chip, which is also BI, and 
from an annulus centred at the galaxy centroid and with an inner and outer 
radii typically $\sim$2.5\arcmin\ and 3.3\arcmin. 
We excluded data from the vicinity
of any point-sources found by the source detection algorithm.
Although the diffuse emission from each galaxy typically had a very low 
surface-brightness on the S1 CCD, we found that using two regions 
in this way with different contributions of source emission enabled the 
background components to be most cleanly disentangled from the source.
The \acis\ focal plane also consists of front-illuminated (FI) chips, 
which have significantly different (and lower) background. 
To obtain an estimate of the background for these chips, 
we extracted spectra from the entirety of each chip, excluding 
detected sources and data towards the edge of the chips where the
exposure-map may be uncertain.

In order to constrain the model, we fitted all
spectra simultaneously, without background subtraction, using \xspec.
Our model  consisted of
a single \apec\ plasma (to take account of the diffuse emission from
the galaxy; the ``source''),
plus background components. These comprised
a power law with $\Gamma=1.41$ (to account for the hard X-ray background),
two \apec\ models with solar abundances and kT$=$ 0.2 and
0.07~keV (to account for the soft X-ray background) and, to model
the instrumental and particle contributions, a broken power law
model and two Gaussian lines with energies 1.7 and 2.1~keV and negligible
intrinsic widths. We have found that this model
can be used to parameterize adequately the template background spectra.
In general, the instrumental contributions of the FI chips were very
similar in shape. Therefore, the background components of all the FI
chips were tied, assuming the normalization scaled with the spectral 
extraction area. For the BI chips, there was some evidence that the 
S1 chip background can be somewhat larger at energies \gtsim 5~keV
(although this is variable).
In order to disentangle the source and background components,
given the general lack of
photons in these spectra, we tied the abundances and
temperatures of the ``source'' \apec\ components between the
extraction regions, but allowed the normalizations to be free.
Where there was a significant improvement in the fit-statistic
if this assumption was relaxed, we allowed the abundances or 
temperatures to fit freely. Notwithstanding, this 
assumption should not significantly affect our results.
This model was able to fit all of the data well.
In our subsequent spectral analysis, we did
not background-subtract the data using the standard templates, but took
into account the background by using appropriately scaled versions
of the models fitted to each CCD, which were added according to the 
overlap between the source region and the CCD. 
We found that the standard background templates fared
much worse than these modelled background estimates when the data
were from regions of low surface-brightness. We discuss the impact of 
the background treatment on our results in \S~\ref{sect_systematics_bkd}.

\section{X-ray images} \label{sect_imaging}
%\clearpage
\begin{figure*}
\centering
\plotone{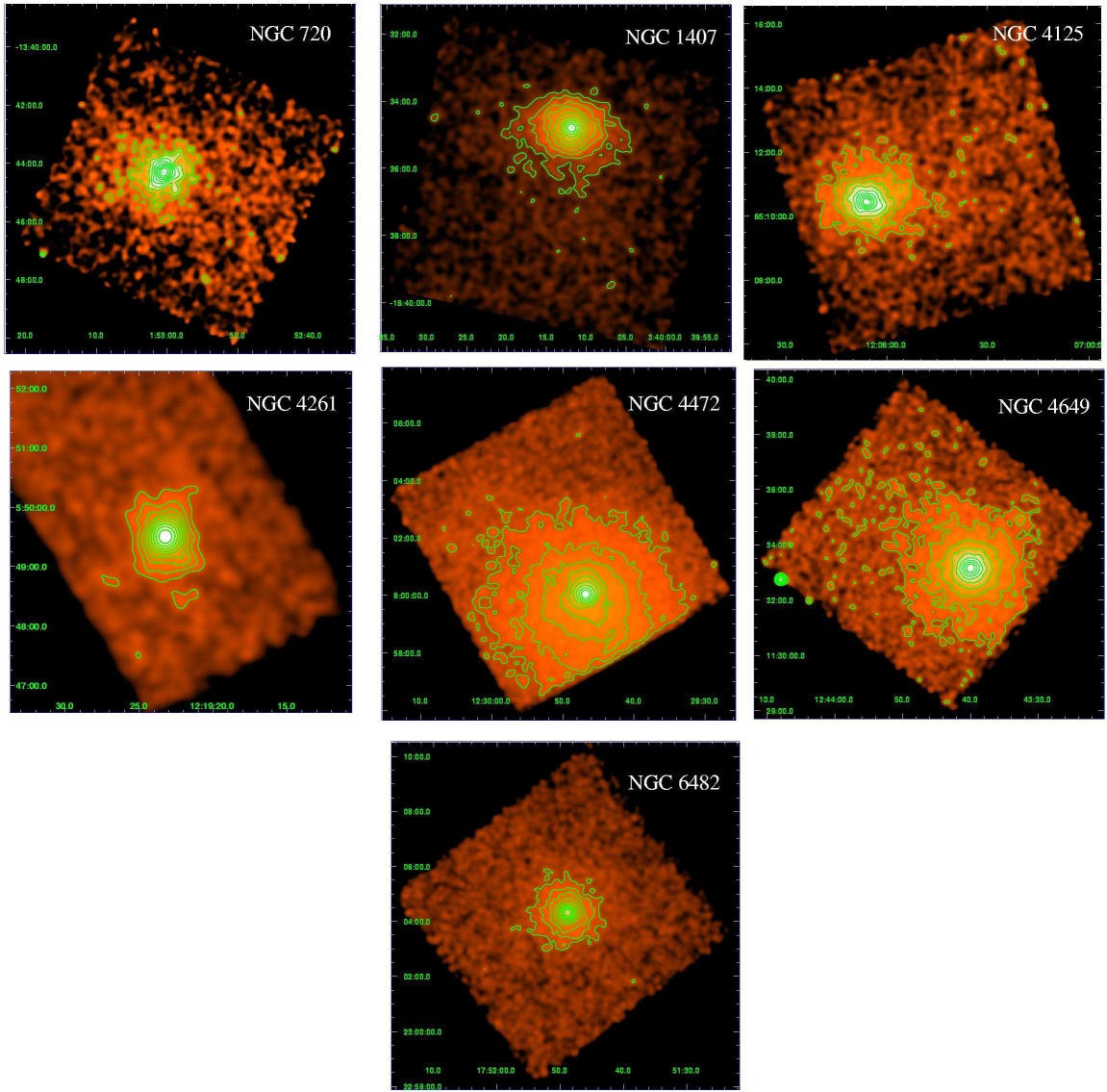}
\caption{X-ray images of each of the galaxies in the sample. None of 
the systems show evidence of large-amplitude disturbances which would
indicate a violation of hydrostatic equilibrium. Some lower-amplitude
asymmetries do persist in some of the images, which we discuss in 
detail in \S~\ref{sect_asymmetry}.}
\label{fig_images}
\end{figure*}
%\clearpage
The X-ray image of each galaxy was examined to identify any 
obvious surface-brightness disturbances or asymmetries 
which would be indicative of clear deviations from hydrostatic equilibrium.
We note that low-level X-ray asymmetries, 
such as the ``fingers of emission'' identified by 
\citet{randall03} in the adaptively-smoothed images of NGC\thin 4649,
probably do not merit concern\footnote{Although the authors suggested these
may arise from bulk convective flow, the spectra do not agree with 
simulations of such.}, as, provided care is taken to avoid
seriously disturbed emission regions, reliable mass profiles can
be inferred even in mildly disturbed systems \citep{buote95a}.

In Fig~\ref{fig_images} we show the 0.1--10.0~keV 
\acis-S3 images of each of the systems. 
These images were first processed to remove point-sources, 
using the \ciao\ tool {\em dmfilth}, which replaces photons in the vicinity
of each point-source with a locally-estimated background.
NGC\thin 4261 contains an AGN which appears as a bright central X-ray
source and there is evidence of a small, low surface-brightness 
jet \citep{zezas05a}. We have also removed these 
sources from the image. The images were flat-fielded with the 
1.7~keV monochromatic exposure-map (although this analysis is insensitive
to the choice of energy), and then smoothed by convolution with a 5\arcsec\
gaussian, to make large-scale structure more apparent. 
Due to the 
low surface-brightness nature of the emission at large radii, it 
is difficult to appreciate X-ray emission outside $\sim$a few arc minutes
in many of the images. However, detailed 
spectral analysis and azimuthally-averaged surface brightness analysis
reveals substantial hot gas extending beyond the edge of the S3 chip
in each system.

None of the objects show very obvious disturbances in their X-ray emission
on the \acis-S3 chip \citep[such as those found in NGC\thin 4636:][]{jones02a}.
Some low-amplitude features are evident such as the faint
jet in NGC\thin 4261 (which is not visible in the above images),
a possible north-south asymmetry in NGC\thin 1407 and some asymmetry,  in particular
an off-axis X-ray enhancement, in NGC\thin 4125. Based on adaptively-smoothed
\xmm\ images, \citet{croston05a} argued that the X-ray emission in 
NGC\thin 4261 is anti-correlated with the galaxy radio lobes.
By inspection of the \xmm\ images, this actually appears to be a very 
low-amplitude effect. It is not obvious in the 
\chandra\ images, although the X-ray isophotes do align somewhat perpendicularly
to the jet. 
In any case, this does not appear to have significantly disturbed
hydrostatic equilibrium, since there is excellent agreement between our 
inferred mass profile and a model comprising stellar plus DM components
(\S~\ref{sect_mass}), which would be an extraordinary coincidence
if hydrostatic equilibrium had been strongly disturbed. 
The limited field-of-view makes it difficult to assess asymmetries and
disturbances on the other chips. NGC\thin 4472 is known, however,
to exhibit a disturbance outside $\sim$6\arcmin\ \citep{irwin96}, but 
mass analysis inside this radius should be reliable. We assess the impact
of all these features in \S~\ref{sect_asymmetry}.

\section{Spectral Analysis} \label{sect_spectra}
Spectral-fitting was carried out in the energy-band 0.5--7.0~keV, 
to avoid calibration uncertainties at lower energies
and to minimize the instrumental background, which dominates at high
energies. The spectra were rebinned to ensure a S/N ratio of at least
3 and a minimum of 20 photons per bin (to validate $\chi^2$ fitting).
We fitted data from all annuli simultaneously using
\xspec. To model the hot gas we adopted a {\bf vapec} component,
plus a bremsstrahlung component for all annuli within the 
twenty-fifth magnitude isophote (\dtwentyfive) of each galaxy, taken
from the Third Reference Catalog of Bright Galaxies 
\citep[RC3:][]{devaucouleurs91}, to
account for undetected point-sources \citep[this model gives a good fit to
the composite spectrum of the detected sources in nearby 
galaxies:][]{irwin03a}.
We used a slightly modified form of the existing \xspec\ {\bf vapec}
implementation so that \zfe\ is determined directly, but for the remaining
elements the abundance ratios (in solar units) were directly
determined with respect to Fe. This was useful since, in general, the data
did not enable us to determine any abundance {\em ratio} gradients and
so we tied the abundance ratios between all annuli. 
Where abundances or abundance ratios could not be constrained, 
they were fixed at the Solar value.
We adopted the solar photospheric
abundances standard of \citet{asplund04a}. We refer the interested 
reader to \citet{humphrey05a} for a detailed discussion of this 
choice and how to convert our results to older abundance
standards. In the interests of 
physically reasonable results, we constrained all abundances and
abundance ratios to the range 0.0--5.0~times solar.
 The absorbing column density (\nh) was fixed at the Galactic value
\citep{dickey90}; the effect of varying \nh\ is discussed in 
\S~\ref{sect_systematics_spectra}.
For NGC\thin 4261, our innermost annulus contained substantial
contamination from the central AGN. However, this was sufficiently
absorbed that the thermal emission from the gas can be clearly
disentangled from it. To account for the AGN emission, we fitted
a highly absorbed (\nh${\rm =10^{+6}_{-4}\times 10^{22} cm^{-2}}$)
power law component ($\Gamma=1.4\pm0.8$). We discuss the impact 
of including this annulus on our fits in \S~\ref{sect_asymmetry}.

To account for projection effects, we used the {\bf projct} model implemented
in \xspec. This model, unfortunately, does not take into account the 
emission from gas outside the outermost shell, which is also projected
into the line-of-sight. To take account of this effect, we assumed that 
the emission outside this shell has the same spectral shape as the 
emission in that shell and a density profile well-described by a $\beta$-model
\citep[\eg][]{buote00c}.
We included an extra spectral component to our fits of each annulus to account
for projected emission from this gas.
To estimate the parameters of the $\beta$-profile, we 
we fitted the galaxy surface brightness, using dedicated software,
in the 0.1--3.0~keV band. Although a single $\beta$-model did not
always match the fine detail of the surface brightness profiles, it
adequately parameterized the data for our purposes (our results are
not expected to be strongly dependent upon the parameters of this fit).

We obtained good fits to the spectra of each galaxy with this model.
The best-fitting abundances 
were in excellent agreement with those of other early-type galaxies
\citep{humphrey05a}, and are shown in Table~\ref{table_abundances}.
We note that \citet{randall05a} found \zsi/\zfe$\simeq$1.7 for
NGC\thin 4649 (adjusting to our abundances standard) when 
fitting the data  from single, large aperture, which they
argued points to substantial enrichment from SN~II, 
in stark contrast to the predominantly SN~Ia enrichment 
we found in such galaxies \citep{humphrey05a}. From our analysis,
however, \zsi/\zfe$\simeq$1, which is more 
consistent with our results for other systems. The discrepancy 
appears to be related to the ``Fe bias'' \citep[where \zfe\ is 
systematically underestimated if one assumes multi-temperature gas
is isothermal:][]{buote00c} which has suppressed their large aperture
\zfe\ in comparison to their spatially-resolved results (which
agree better with our measurement).

Error-bars were computed {\em via} the Monte-Carlo technique which
we have extensively used in past analyses \citep[\eg][]{buote03a}.
We simulated spectra from the best-fit models, which were then
fitted exactly analogously to the real data. We performed 25
simulations, which were sufficient to assess the distribution
of the fit parameters about the best-fit values; the standard
deviation of this distribution corresponds to the 1-$\sigma$
confidence region.Assuming that we have found the global minimum, 
and the fit statistic follows a $\chi^2$ distribution this is 
statistically equivalent to searching the parameter space for 
changes in the fit statistic.
Temperature and density profiles are discussed below 
(\S~\ref{sect_temp_profiles} and \S~\ref{sect_mass_results})
%\clearpage
\begin{deluxetable*}{lrrrrrrrr}
\tabletypesize{\scriptsize}
\tablecaption{Emission-weighted average abundances\label{table_abundances}}
\tablehead{
\colhead{Galaxy} & \colhead{$\chi^2$/dof} & \colhead{\zfe} & \colhead{\zo/\zfe} & \colhead{\zne/\zfe} & 
\colhead{\zmg/\zfe} & \colhead{\zsi/\zfe} & \colhead{\zs/\zfe} & \colhead{\zni/\zfe} 
}
\startdata
NGC\thin 720$^1$  & 383.4/357& $0.80^{+0.45}_{-0.24}$ & 0.30$\pm 0.28$ & 0.68$\pm0.67$ & 1.26$\pm0.35$ & \ldots& \ldots & \ldots \\
NGC\thin 1407$^1$ & 222/221&  2.1$^{+1.1}_{-0.9}$$^\dagger$& 0.37$^{+0.21}_{-0.25}$ & \ldots & 1.10$\pm 0.23$ & 1.21$^{+0.31}_{-0.27}$ & 2.2$\pm1.1$ & 3.3$^{+1.7}_{-1.3}$\\
NGC\thin 4125 & 327/307 & 0.55$^{+0.22}_{-0.13}$ & 0.29$^{+0.13}_{-0.09}$ & 0.62$\pm$0.14 & 0.33$\pm0.20$ & \ldots & \ldots & \ldots \\
NGC\thin 4261 & 307/319 & 1.72$\pm0.50^\dagger$ & $<$0.23 & 0.36$^{+0.79}_{-0.36}$ & 0.83$\pm$0.23& 1.2$\pm$0.4& \ldots & 1.8$^{+2.3}_{-1.8}$ \\
NGC\thin 4649 & 563/491 & 2.32$^{+0.87}_{-0.37}$ & $<0.15$ & \ldots & 0.97$\pm$0.13 & 1.02$\pm$0.13 & \ldots & 1.42$^{+0.85}_{-0.73}$ \\
NGC\thin 4472$^1$&  785/740 & $1.4^{+1.7}_{-0.4}$$^\dagger$& 0.51$\pm$0.12 & 0.95 $\pm0.44$ & 1.02$\pm$0.11 & 1.25$\pm$0.11 & 2.36$\pm$0.33 & 3.28$\pm0.61$\\
NGC\thin 6482    & 256/262 & $>$2.5 & 0.34$\pm$0.20 & \ldots & 1.15$\pm$0.18 & 1.3$\pm$0.3 & \ldots & 3.2$^{+1.5}_{-1.2}$
\enddata
\tablecomments{The best-fitting globally-averaged emission-weighted abundances
 and abundance ratios  for each galaxy, shown along with the quality of fit. 
Statistical errors represent the 90\% confidence region.
Where we were able to constrain an abundance gradient, we estimated an 
emission-weighted \zfe, extrapolated
over a large aperture \citep[see][]{humphrey05a}; 
those affected galaxies are marked ($^\dagger$).
$^1$---results taken from \citet{humphrey05a}.
Where parameters could not be constrained, they were fixed
at the Solar value, and listed as ``\ldots''.}
\end{deluxetable*}
%\clearpage

\begin{figure*}
\centering
\plotone{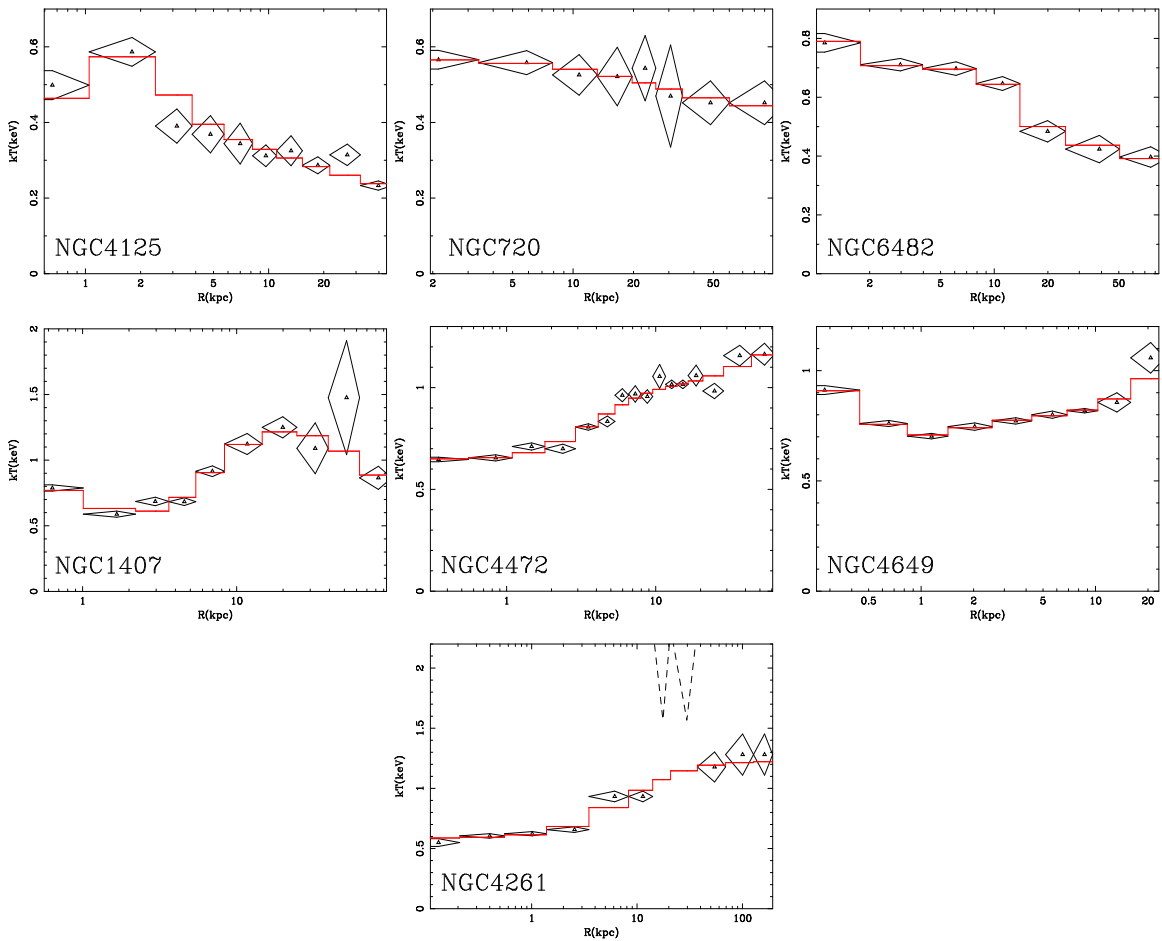}
\caption{Deprojected temperature profiles of each galaxy, ordered by 
\mvir\ (Table~\ref{table_syserr}). The data-points are shown, along
with the best-fit parameterized model determined from simultaneously fitting
the \rhog\ and temperature profiles (see text). Where data-points were excluded
from the fitting, they are denoted by dashed lines. Errors shown are 1-$\sigma$.}
\label{fig_temp}
\end{figure*}

\begin{figure*}
\centering
\plotone{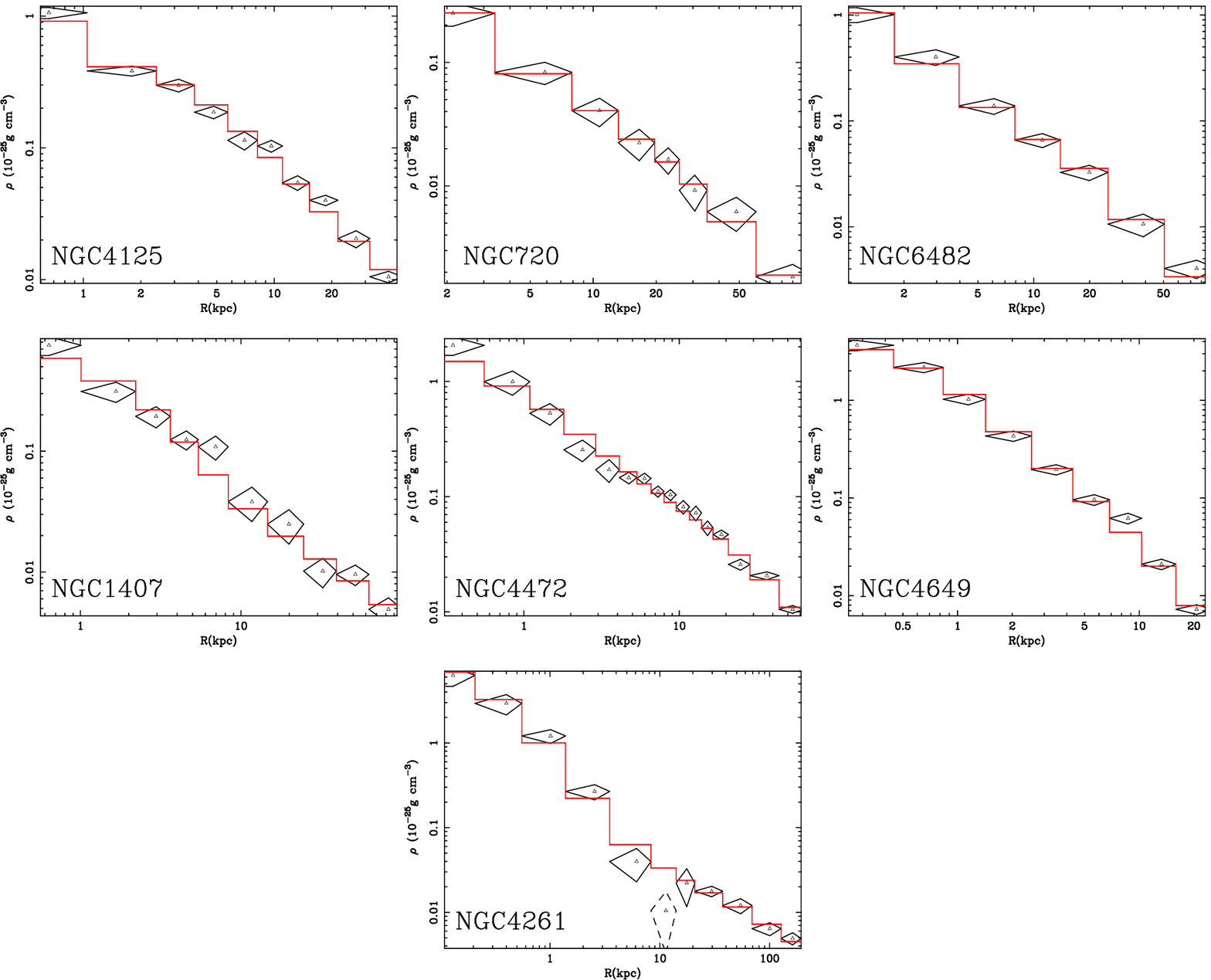}
\caption{Deprojected density-profiles of each galaxy, shown with the best-fit
AC NFW+stars model from the ``assumed potential'' modelling (\S~\ref{sect_potential}).
Data-points excluded from the fit are indicated by dashed lines. 
Errors shown are 1-$\sigma$.}
\label{fig_density}
\end{figure*}
%\clearpage
\section{Mass modelling} \label{sect_mass}

\subsection{Assumed potential method} \label{sect_potential}
We adopted two complementary approaches in order to determine the 
mass profiles of the galaxies
in the sample. The first method, discussed here, was found to be 
less sensitive to the assumptions of the modelling and therefore was
adopted as our default. We discuss our alternative approach in
 \S~\ref{sect_xmass}. 

Starting with a parameterised 
model for the temperature (T) and gravitating mass ($M_{grav}$) 
profiles, the equation of 
hydrostatic equilibrium can be solved for \rhog\ thus:
\begin{equation} 
\ln \left( \frac{\rho_g}{\rho_{g0}} \right) = - \ln \left( \frac{T}{T_0} \right)
- G \mu m_p \int^R_{R_0} \frac{M_{grav}(<R)}{kT R^2} dR
\label{eqn_hydrostatic_rho}
\end{equation}
where R is the radius from the centre of the gravitational potential,
\rhog\ is the gas density, $\rho_{g0}$ and $T_0$ are density and temperature
at some ``reference'' radius $R_0$,
k is Boltzmann's constant, G is the universal gravitational constant,
$m_p$ is the atomic mass unit and $\mu$ is the mean atomic weight of the gas.
In our fitting we explicitly ignored the contribution of the gas to
the gravitating mass, but we subsequently verified this contributed
\ltsim 1\% of the total gravitating matter within 100~kpc, justifying
this assumption.
We developed software to fit \rhog\ and temperature
profiles simultaneously using this procedure. 
For speed we assumed that the density and temperature data-points were 
each evaluated at a single point, the radius of which was given by:
\begin{equation}
\bar{R_i} = \left( 0.5*(Rin_i^{1.5}+Rout_i^{1.5}) \right)^{2/3} \label{eqn_radius}
\end{equation}
where $Rin_i$ and $Rout_i$ were the inner and outer radius of the bin
\citep[see][]{lewis03a}.

\subsection{Temperature profiles} \label{sect_temp_profiles}
There were considerable differences
in the temperature profiles from object to object
(Fig~\ref{fig_temp}), so that we were not
able to adopt a ``universal'' profile for all of the systems.
{\em A priori} we do not expect any particular form for the temperature
profile and so we determined appropriate functional forms for our 
temperature models empirically. Based on experience, the following
``toolbox'' of models provided adequate flexibility to ensure
at least one model can describe
the temperature profiles reasonably well \citep[see][]{buote06a}:
\begin{eqnarray}
T & = & T_0 + T_1 \left[1+x^{-\epsilon}\right]^{-1} \label{eqn_trise2}\\
T & = & \left[T_0 + T_1 x^{p_1}\right]e^{-x^{p_e}} + T_2 x^{p_2} \left[1 - e^{-x^{p_e}}\right]
 \label{eqn_pow2expcut2} \\
T & = & \frac{A}{A+B}\left[ T_0+T_1\left( \frac{x_1}{1+x_1}\right)^{p_1} \right] 
+ \nonumber\\ & & \frac{B}{A+B}\left[ T_2+T_3(1+x_2)^{-p_2} \right] \label{eqn_twophase2}
\end{eqnarray}
where $x=(r/r_c)$,  $x_1=(r/r_{c1})$, $x_2=(r/r_{c2})$,
$A=(1+r/r_{t1})^{-3\beta_1}$ and $B=\epsilon (1+r/r_{t2})^{-3\beta_2}$.
$T_0$, $T_1$, $T_2$, ${T_3}$, $r_c$, $r_{c1}$, $r_{c2}$,
$r_{t1}$, $r_{t2}$,  $p_1$, ${p_2}$, $p_e$ and $\epsilon$ are parameters
of the fit. 
For NGC\thin 4261, we ignored temperature data-points from 15--25~kpc,
which were poorly-determined and seemed erroneous. We experimented with
fitting the projected (rather than deprojected) spectra, and found no
evidence of any features (in either temperature or density) around
this range of radii, strongly implying that they
arise solely due to deprojection noise. The temperature profiles and best-fit 
models are shown in Fig~\ref{fig_temp}.

Our deprojected temperature profiles generally agree with those 
appearing in the literature for these objects. (Although most
of these are projected profiles, typically deprojection does not 
strongly alter the overall character of the temperature profile.)
\citet{osullivan03a}
reported \rosat\ profiles for all of the galaxies which, although
substantially less well-constrained, agree well with our results. 
Likewise our NGC\thin 4472 temperature profile agrees well with 
the (less well-constrained) \rosat\ profile of \citet{irwin96}.
Our profile for NGC\thin 4649 is in reasonable agreement with the 
projected  \xmm\ measurements of \citet{randall05a}, and likewise our
measured profile of NGC\thin 6482 agrees with the deprojected
results of \citet{khosroshahi04a}. Our temperature profiles for 
NGC\thin 1407, NGC\thin 720 and NGC\thin 4472 were also in agreement
with those we reported in \citet{humphrey05a}.

\subsection{Mass-fitting results} \label{sect_mass_results}
%\clearpage
\begin{figure*}
\centering
\plotone{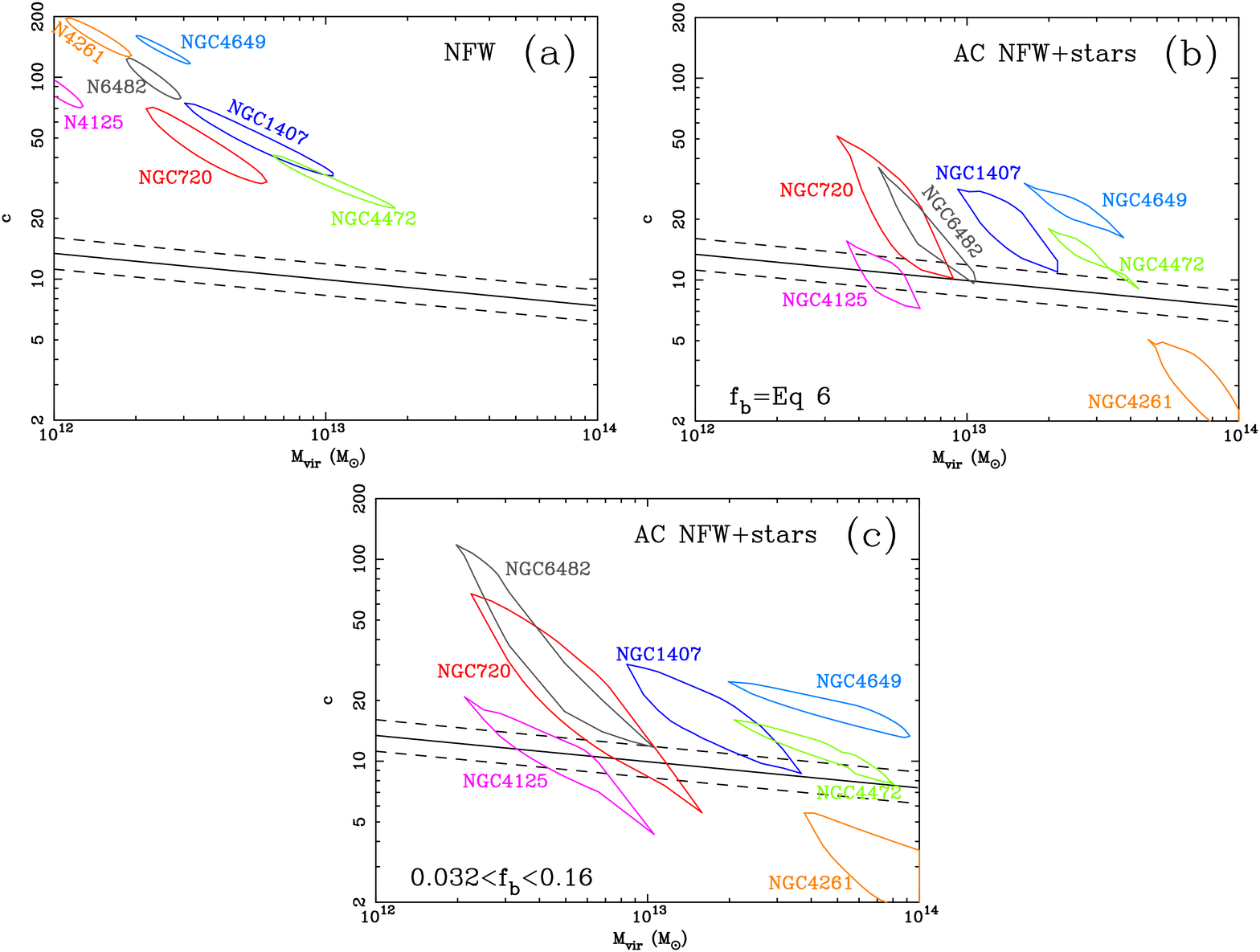}
\caption{1-$\sigma$ (\mvir,c) confidence regions for each object. 
Panel (a) shows the confidence contours found when fitting the  NFW 
model. In panel
(b) we show the contours for the AC NFW+stars model, restricting
\fbaryons\ by Eq~\ref{eqn_baryons} (see text), and panel (c) shows
contours for the same model, but now restricting 
0.032$\leq$\fbaryons$\leq$0.16. The results for the NFW+stars model
were very similar to those for the AC NFW+stars model. On each plot, 
we also show the prediction from the ``toy model''
of \citet{bullock01a} (solid line) and the approximate 
1-$\sigma$ scatter in the simulated DM halos (the region bounded by 
the dotted lines).}
\label{fig_confidence}
\end{figure*}
%\clearpage
\begin{deluxetable*}{llll}
\tablecaption{Quality of the mass fits\label{table_chisq}}
\tabletypesize{\scriptsize}
\tablehead{ \colhead{Galaxy} & \colhead{NFW} & \colhead{NFW+stars} & \colhead{AC NFW+stars}
}
\startdata
NGC\thin 720  & 1.9/9& 1.0/8& 0.9/8\\
NGC\thin 1407 & 26.7/9& 20.7/8& 20.7/8\\
NGC\thin 4125 & 23.4/11 & 9.5/10 & 10.8/10 \\
NGC\thin 4261 & 22.6/12 & 14.0/11 & 14.0/11 \\
NGC\thin 4472 & 35.2/20 & 34.9/20 & 35.2/20 \\
NGC\thin 4649 & 30.2/7 & 11.0/6 & 11.5/6 \\
NGC\thin 6482 & 0.5/5 & 2.4/4 & 1.7/4
\enddata
\tablecomments{The $\chi^2$/dof of the fits to the density and
temperature profiles used to infer the mass, for the three 
basic mass-models adopted. For the NFW+stars and AC NFW+stars models,
we constrain \fbaryons\ to Eq~\ref{eqn_baryons}.}
\end{deluxetable*}

\begin{deluxetable*}{lllll|rrrr}
\tablecaption{Best-fitting NFW+stars results\label{table_results}}
\tabletypesize{\scriptsize}
\tablehead{ \colhead{Galaxy} & \multicolumn{4}{l}{\fbaryons=Eq~\ref{eqn_baryons}} & \multicolumn{4}{l}{0.032$\leq$\fbaryons$\leq$0.16}\\ 
\colhead{} & \colhead{\mvir ($10^{12}$\msun)} & \colhead{\rvir (kpc)} & \colhead{c} & \colhead{\fbaryons} &\colhead{\mvir ($10^{12}$\msun)} & \colhead{\rvir (kpc)} & \colhead{c} & \colhead{\fbaryons} }
\startdata
NGC720 &$6.6^{+2.4}_{-3.0}$ &$480^{+50}_{-90}$ &$18.^{+30.}_{-8.}$ &$0.044^{+0.037}_{-0.003}$ &$6.6^{+6.0}_{-4.3}$ &$480\pm 120$ &$18.^{+49.}_{-10.}$ &$0.044^{+0.095}_{-0.012}$ \\
NGC1407 &$16.\pm 6.$ &$650^{+80}_{-100}$ &$18.^{+11.}_{-7.}$ &$0.065^{+0.041}_{-0.001}$ &$21.\pm 15.$ &$720^{+140}_{-200}$ &$15.^{+16.}_{-6.}$ &$0.032^{+0.130}_{-0.001}$ \\
NGC4125 &$6.2^{+0.8}_{-2.3}$ &$470^{+20}_{-70}$ &$10.^{+5.}_{-2.}$ &$0.039^{+0.035}_{-0.001}$ &$7.2^{+1.4}_{-4.9}$ &$500^{+30}_{-160}$ &$9.3^{+11.}_{-2.1}$ &$0.032^{+0.13}_{-0.001}$ \\
NGC4261 &$67.^{+41.}_{-15.}$ &$1040^{+200}_{-90}$ &$3.7\pm 1.7$ &$0.14^{+0.01}_{-0.03}$ &$57.^{+260}_{-15.}$ &$990^{+760}_{-100}$ &$4.0\pm 2.0$ &$0.16^{+0.00}_{-0.13}$ \\
NGC4472 &$33.^{+6.}_{-10.}$ &$820^{+50}_{-100}$ &$13.^{+4.}_{-2.}$ &$0.084^{+0.037}_{-0.001}$ &$63.^{+17.}_{-44.}$ &$1020^{+90}_{-300}$ &$10.0^{+7.}_{-2.}$ &$0.032^{+0.13}_{-0.00}$ \\
NGC4649 &$35.^{+7.}_{-13.}$ &$840^{+60}_{-120}$ &$21.^{+6.}_{-3.}$ &$0.086^{+0.037}_{-0.001}$ &$93.^{+26.}_{-73.}$ &$1200^{+100}_{-500}$ &$15.^{+11.}_{-3.}$ &$0.032^{+0.12}_{-0.00}$ \\
NGC6482 &$7.1^{+4.4}_{-1.7}$ &$500^{+90}_{-40}$ &$18.^{+13.}_{-8.}$ &$0.075^{+0.013}_{-0.032}$ &$3.6^{+5.5}_{-1.5}$ &$390^{+140}_{-70}$ &$38.^{+76.}_{-24.}$ &$0.16^{+0.00}_{-0.10}$ 
\enddata
\tablecomments{The best-fitting results for the NFW+stars model. All
error-bars shown correspond to 90\% confidence regions. The
fit results for the AC NFW+stars model are very similar, and are 
shown in Fig~\ref{fig_confidence}. Results are shown for the fits
using the two different constraints on \fbaryons\ we adopted (see text).}
\end{deluxetable*}
%\clearpage
We tested three different mass-models against the data. 
In order to investigate the suggestion that historically large
c values found based on X-ray analysis were an artefact of the 
omission of the stellar mass, as well as to investigate the 
scenario of \citet{loeb03a}, we first tested a model comprising a 
single NFW profile. Although stellar kinematical results would
seem to rule out the \citeauthor{loeb03a} picture, our analysis
of more massive systems \citep{gastaldello06a} does suggest that the
stellar mass may not be uniformly required in all systems.  
In order to take into account the stellar mass, we fitted a 
model comprising an NFW DM component, plus
a \citet[][hereafter H90]{hernquist90} mass component, the \reff\ 
of which being fixed to that measured in the \ks-band 
(Table~\ref{table_obs}). The H90 model is, in projection, a good
approximation to the familiar de Vaucouleurs profile of elliptical
galaxies.  To test whether the DM halos retained any
evidence of their response to baryonic
condensation, we further adopted an H90 component, plus an NFW component
modified by the adiabatic contraction model of \citet{gnedin04a}\footnote{Available publicly from \href{http://www.astronomy.ohio-state.edu/$\sim$ognedin/contra/}{http://www.astronomy.ohio}-state.edu/$\sim$ognedin/contra/}.
Hereafter, we refer to these three models as, respectively, NFW,
NFW+stars and AC NFW+stars. Our computed \mvir\ for each system included both
dark and stellar mass. For NGC\thin 4472 and NGC\thin 4649, which lie
in Virgo, there is the possibility that the DM halo may have experienced
some tidal truncation at a radius $<$\rvir. Our measured Virial quantities
relate to the original halo prior to truncation.
We also experimented with replacing the NFW component with the less
cuspy  \citet[][hereafter N04]{navarro04a} model, which gives 
an improved fit to DM halos in high-resolution N-body simulations.
However, since the \mvir-c relation was calibrated using
the NFW model we treat this choice as a systematic effect and it is
discussed in \S~\ref{sect_n04}.
For NGC\thin 4261 we ignored a deviant \rhog\ data-point at $\sim$11~kpc, 
in addition to the excluded temperature data-points discussed above. %DONE

In Fig~\ref{fig_density} we show the density profiles (along with the 
best-fitting AC NFW+stars model, which is described below).
In Fig~\ref{fig_confidence}(a) we show the best-fitting 
1-$\sigma$ contours of c {\em versus} 
\mvir\ for the NFW model fitted to each galaxy. The fits were 
typically, but not uniformly, poor (Table~\ref{table_chisq}).
We found very large ($\gg$20) values for c, completely 
inconsistent with the expectation of N-body simulations.

Good constraints on the global halo properties typically require interesting 
density and temperature constraints over as large a radial range as possible.
In our case, the absence of data outside $\sim$50--100~kpc ($\sim$0.1--0.2\rvir) 
therefore
makes the inner data-points critical in determining the profile of the 
halo. Unfortunately, since the scale radius of a galaxy-size DM halo is 
$\sim$10--30~kpc, there is some degeneracy between the DM and 
stellar mass components at small radii.
As we discuss in \S~\ref{sect_mass_to_light} there are considerable
uncertainties in estimating a reliable mass-to-light (M/L) ratio from
the characteristics of the stellar population. We found that the 
results are extremely sensitive to the stellar M/L adopted; we found
that varying this ratio by as little as 20\% could cause \mvir\
variations of $\sim$50--100\% \citep[see][]{humphrey05b}.
It was therefore necessary to allow the stellar mass to be determined as a 
parameter of the fit. This, unfortunately, made it very difficult
to constrain \mvir\ or c, unless additional constraints were applied
to the fit.

One way to achieve this is to constrain the fit to lie on the mean
\mvir-c relation predicted from N-body simulations \citep[\eg][]{bullock01a}.
Although this would prevent our measuring \mvir\ and c independently,
it would enable us to determine whether the 
data are consistent with the mean relation.
However, this relation was determined for an unbiased sample of 
DM halos, whereas our selection criteria (\S~\ref{sect_targets}) should 
bias us towards systems which have not recently had a merger (implying
earlier-forming, hence more concentrated, objects). 
Furthermore, individual halos are not expected
to lie exactly on the mean \mvir-c relation, but be scattered
about it. Nevertheless, we experimented with applying this constraint.
The data for each galaxy were consistent with this model, but
we found \mvir\ was generally poorly constrained, and extremely sensitive
to any scatter we introduced about the mean \mvir-c relation.

A far more useful way to constrain the fit was to restrict the
total baryon fraction (\fbaryons) in the system.
Such a constraint is useful since we found that, for a given system,
\fbaryons\ determined from our fits was strongly anti-correlated with
the measured \mvir. To estimate \fbaryons, we 
computed the gas mass by extrapolating our \rhog\ model from the 
centre of the innermost radial bin to the Virial radius.
The contribution of stars to the total baryon fraction was derived
from the stellar mass found by our fits. 
We crudely took into account the fact that 
not all of the stellar mass within \rvir\ is necessarily 
contained in the central galaxy by scaling this mass by the ratio 
of the total B-band light of all putative ``group'' members listed
in the catalogue of  \citet[][hereafter G93]{garcia93} to that of the central galaxy.
This is likely to overestimate slightly the stellar mass content, since
it assumes the same stellar M/L ratio for all low-mass companions/ group
members, 
whereas some fraction of these are likely to have substantial young stellar
populations, with lower M/L ratios. 
We discuss the impact of this assumption in \S~\ref{sect_systematics_baryon_fraction}.
G93 lists NGC\thin 4649 as belonging to the NGC\thin 4472
``group'', whereas they both have distinct X-ray halos, indicating
they are, in fact, distinct systems. As a zeroth order approximation, we 
therefore assumed that the total B-band luminosity was divided between
the two ``subgroups'' in proportion to the central galaxy's B-band 
luminosity. In practice, between $\sim$25\% (for NGC\thin 4472) and
84\% (for NGC\thin 4125) of the B-band light of the system resides in the 
central galaxy. NGC\thin 6482 was not listed in G93,
but as it is known to be relatively isolated \citep{khosroshahi04a},
we assumed that $\sim$80\% of its mass is in the central galaxy, consistent
with the other relatively isolated systems.

Based on hydrodynamical simulations incorporating gas cooling and 
supernovae feedback, \citet{kay03a} predicted \fbaryons\ as a function
of Virial temperature for systems with \mvir\gtsim a few $\times 10^{12}$\msun.
Fitting their data by eye, converting from Virial temperature to mass,
and assuming a Universal baryon fraction of 0.16, we estimate
\begin{eqnarray}
f_{b} = & 0.062 \log_{10}(M_{14})+0.13 & (M_{14}<1.02) \label{eqn_baryons} \\
f_{b} = & 0.14                         & (M_{14}>1.02)\nonumber 
\end{eqnarray}
where ${M_{14}=M_{vir}/10^{14} M_\odot}$, with an approximate scatter of 
$\pm$0.02. 
Adopting this constraint (including the allowed range of scatter)
and fitting the NFW+stars model resulted
in significant improvements in the fit quality over the simple NFW 
model (Table~\ref{table_chisq}).
We show  the resulting c-{\em versus}-\mvir\ contours in 
Fig~\ref{fig_confidence}(b), and 
summarise our results in Table~\ref{table_results}. 
Clearly adding the stellar mass component
allows the DM halos to be substantially less concentrated, since
less DM is required in the centre of the halo.
These results were  in much better agreement with the results
of N-body simulations than those obtained with the NFW model. 
There is, however, a slight trend towards 
more concentrated halos than \lcdm. 
%\clearpage
\begin{figure*}
\centering
\plotone{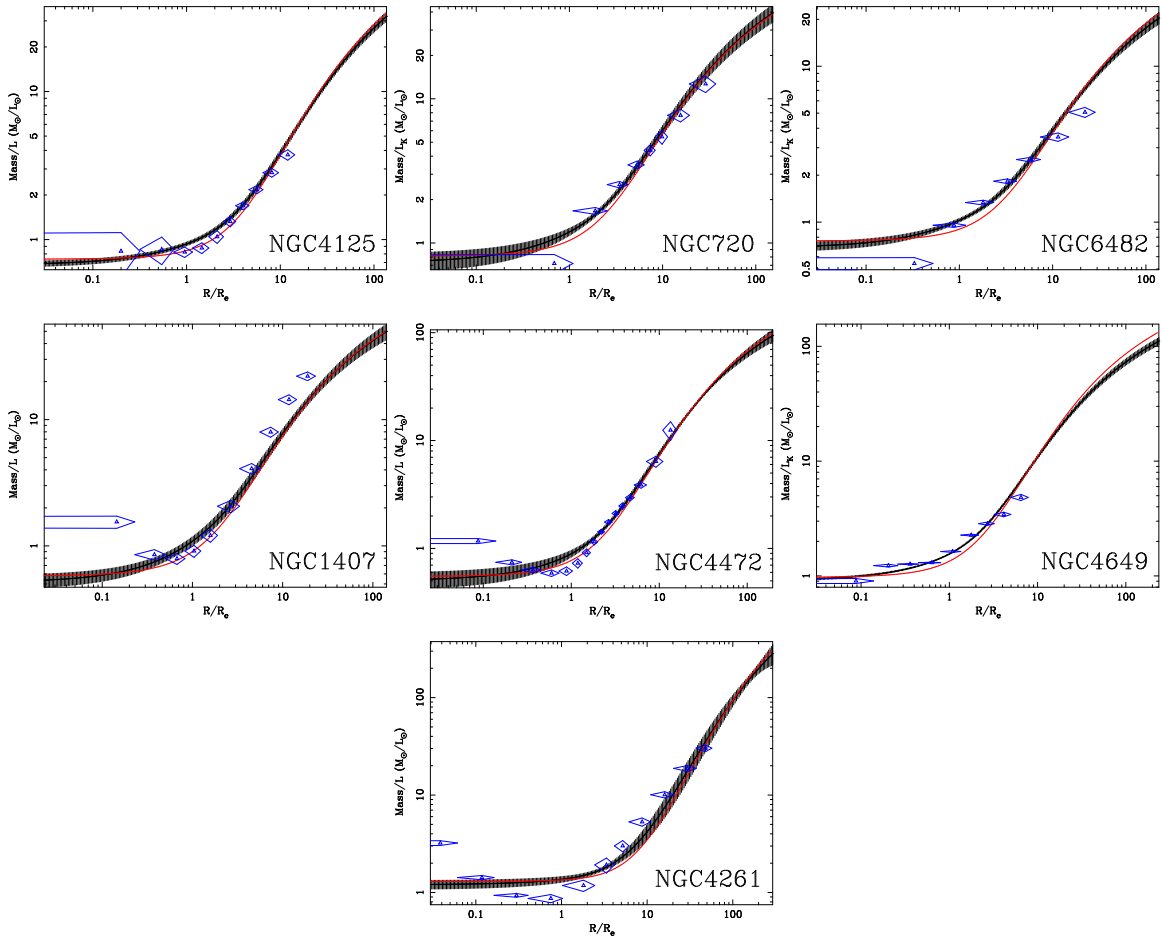}
\caption{The K-band mass-to-light profile of each galaxy. Results are shown
both for the NFW+stars and the AC NFW+stars models. For the latter the shaded 
region indicates the approximate 1-$\sigma$ confidence region.
We also show data-points derived from the ``parameterized profile''
mass-modelling described in \S~\ref{sect_xmass}. We stress that these
data-points are {\em not} fitted by the models shown here, but they
are derived independently.}
\label{fig_mass_to_light}
\end{figure*}
%\clearpage

Since the results of simulations can be sensitive to the rather uncertain
process of feedback, we additionally adopted, as a 
somewhat less restrictive constraint on \fbaryons, 0.03$<$\fbaryons$<0.16$,
the lower limit being $\sim$the lowest values found in
\citeauthor{kay03a}'s simulations. 
The shallower potential of very low-mass halos
makes it more difficult for them to hold onto their hot gas,
and so our lower limit on \fbaryons\ may be an overestimate if 
the Virial mass is small. However, imposing a lower limit on
\fbaryons\ in the fitting algorithm actually works to exclude
the most massive solutions, which we would expect to be closer 
to baryonic closure \citep{mathews05a}. It is conceivable that some 
more massive (\mvir\gtsim $10^{13}$\msun) systems are rare examples 
of ``dark'' groups which have unusually low baryon fractions,
a possibility we return to in \S~\ref{sect_ngc1407}.
Notwithstanding, the results are shown in Figs~\ref{fig_confidence}(c), 
which are qualitatively similar to those obtained applying the more 
restrictive constraint on \fbaryons, although with larger 
uncertainty.

In Fig~\ref{fig_mass_to_light}
we show the gravitating mass to K-band light (\mgrav/\lk) ratio profile
implied by our best-fit models for each system.
In each case we found that the X-ray emission was considerably
more extended than the optical light.
We also show data-points estimated from ``parameterized profile''
mass modelling (\S~\ref{sect_xmass}),
which tend to agree reasonably well; the slight systematic 
differences between the profiles are an artefact of the assumptions used
to derive the data-points and we discuss this in detail in \S~\ref{sect_xmass}.
Clearly \mgrav/\lk\ increases
very slowly with radius within \reff, rising very steeply outside this 
range. This arises naturally from the very different shapes of the stellar
and DM halos, and is similar to M/L profiles seen from stellar 
kinematics and the results of  \citet{brighenti97a} for NGC\thin 4472
and NGC\thin 4649.
By \rvir, \mgrav/\lk\ reaches as high as $\sim$20--40 
\msun/\lsun\ for the galaxy-scale systems or 
$\sim$100-200 \msun/\lsun\ for the group-like objects. We stress
that this only includes the light of the central galaxy which,
for the group-like systems may be a little as $\sim$25\% of the total
luminosity.

\subsection{Parameterized profile mass modelling} \label{sect_xmass}
%\clearpage
\begin{figure*}
\centering
\plotone{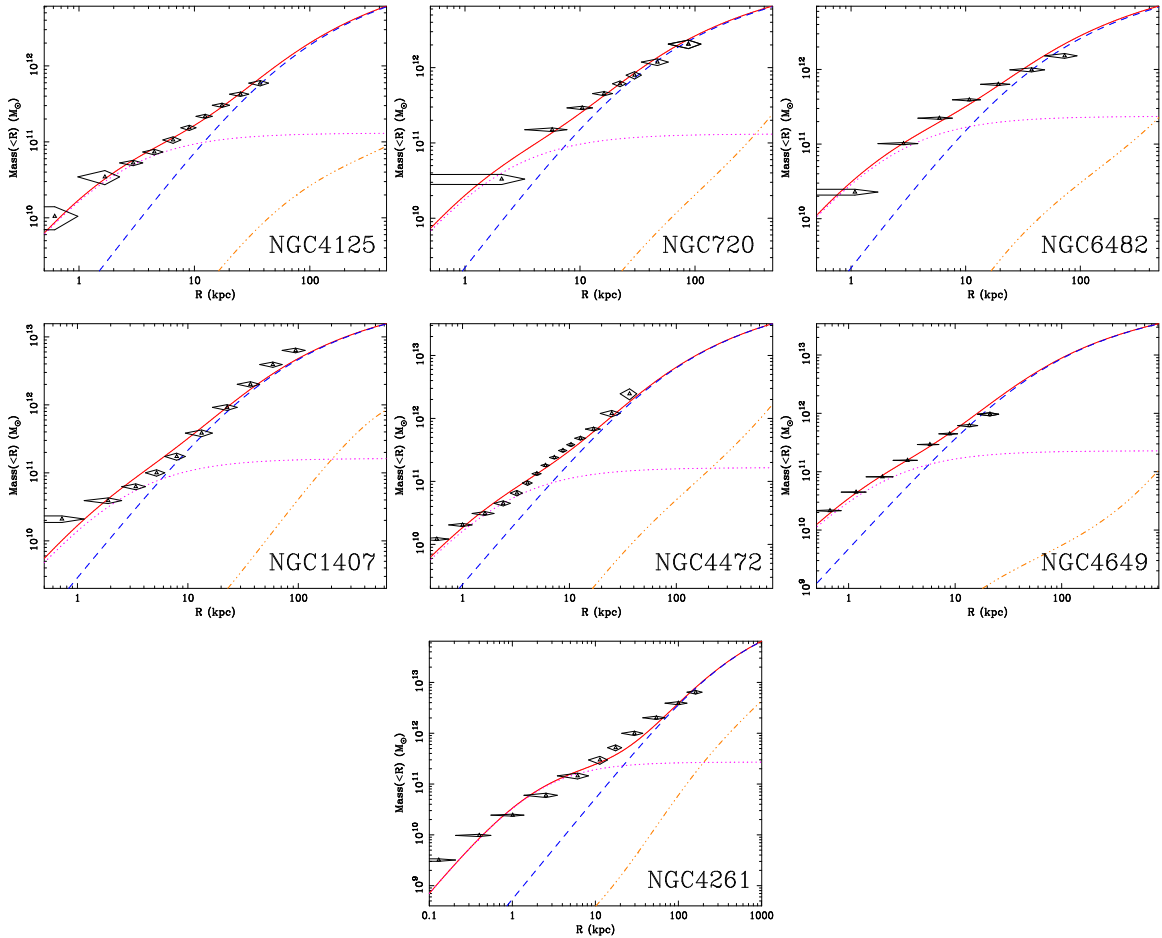}
\caption{Mass profiles for each galaxy. The data-points were computed
using parameterized profile modelling (\S~\ref{sect_xmass}).
In addition we show the best-fit NFW+stars mass models from assumed potential
modelling, which generally agree reasonably well with the data-points,
indicating the consistency of both approaches to determine the mass
profiles. We show the total gravitating mass model (solid line)
and, separately, the stellar mass contribution (dotted line),
the DM contribution (dashed line) and the gas mass (dash-dot-dot line). 
The models are extrapolated
out to \rvir.Errors shown are 1-$\sigma$.}
\label{fig_mass}
\end{figure*}
%\clearpage
We briefly discuss here an alternative technique to determine the 
mass profiles of X-ray bright objects which we have extensively
employed in our previous studies, as well as the companion
papers to this present work 
\citep[\eg][]{lewis03a,buote06a,gastaldello06a,zappacosta06a}.
This technique, which we here dub the ``parameterized profile'' method
involves parameterizing independently the temperature and density 
profiles of the system with simple, empirical models. These functions
were then inserted into the equation of hydrostatic equilibrium,
which we solved for the mass enclosed within any given radius.
The temperature profiles were parameterized with the models
discussed in \S~\ref{sect_temp_profiles}, and to fit \rhog\ we 
adopted, where appropriate a, $\beta$-model, a ``double-$\beta$'' model
or a ``cusped-$\beta$'' model, defined, respectively, as:
\begin{eqnarray}
\rho_g & = & \rho_{g0} \left[ 1+ (r/r_c)^2 \right]^{-3\beta /2} \label{eqn_beta}\\
\rho_g & = & \sqrt{\rho_{g0} \left[ 1+ (r/r_c)^2 \right]^{-3\beta} + \rho_{g1} \left[ 1+ (r/r_{c2})^2 \right]^{-3\beta_2}} \label{eqn_twobeta} \\
\rho_g & = & \rho_{g0} 2^{3\beta/2-\epsilon/2}(r/r_c)^{-\epsilon}\left[ 1+ (r/r_c)^2 \right]^{-3\beta /2+\epsilon/2} \label{eqn_cusp}
\end{eqnarray}
Where the parameters $\rho_{g0}$, $\rho_{g1}$, $r_c$, $r_{c2}$,
$\beta$, $\beta_2$ and $\epsilon$ are determined by the fit.
Fitting these models to the simulated temperature and density profiles 
discussed in \S~\ref{sect_spectra} (which were used therein to estimate the 
error-bars on kT and \rhog\ in each data-bin) allowed us to estimate the
scatter in the mass data-points arising from statistical noise, and
hence the error-bars. %DONE

For a full discussion of this technique, we refer the interested
reader to \citet{buote06a}, who demonstrate the good agreement 
typically found between this method and the assumed potential
modelling of \S~\ref{sect_potential}, when fitting high-quality data.
However, the mass data-points, especially at the innermost and outermost
radii, are rather sensitive to the parameterized models adopted to fit
the temperature and, especially, \rhog. The systematic uncertainty
introduced by the choice of \rhog\ model can be considerably  
larger than the statistical error. For our purposes the absence of data 
at very large radii, which are vital to constrain the curvature of the mass 
model, exacerbated by the uncertainty introduced at small radii due to
the uncertain stellar mass-to-light ratio, magnified the impact of these 
systematic effects. Notwithstanding these reservations, it is still interesting
to compare the results obtained {\em via} both mass-fitting methods. 
We show in Fig~\ref{fig_mass} the mass data-points computed using parameterized
potential modelling, along with the best-fitting mass models
found in \S~\ref{sect_potential}. Clearly there is good overall agreement 
between the two methods although there are some systematic differences,
which reflect the systematics inherent in our choice of 
parameterized model for \rhog.

\subsection{Stellar mass-to-Light ratios} \label{sect_mass_to_light}
%\clearpage
\begin{deluxetable*}{llll|rr}
\tablecolumns{5}
\tablecaption{Stellar mass-to-light ratios\label{table_mass_to_light}}
\tabletypesize{\scriptsize}
\tablehead{ \colhead{Galaxy} & \colhead{\lk/\lb} &
    \multicolumn{2}{c}{Fitted \mstars/\lk\ (\msun/\lsun)}  & 
    \multicolumn{2}{c}{Pop.\ synthesis \mstars/\lk\ (\msun/\lsun)}\\
    \colhead{} & \colhead{} & \colhead{NFW+stars} &  \colhead{AC NFW+stars} & 
    \colhead{Salpeter IMF} & \colhead{Kroupa IMF}
}
\startdata
NGC\thin720     & 5.5 & 0.77$^{+0.52}_{-0.71}$& 0.54$^{+0.42}_{-0.48}$& 0.54$\pm$0.11& 0.35$\pm$0.07\\
NGC\thin 1407   & 4.8 & 0.52$^{+0.25}_{-0.32}$& 0.35$\pm$0.25& 1.6$\pm$0.2& 1.1$\pm$0.1\\
NGC\thin 4125   & 3.8 & 0.72$\pm$0.11& 0.53$\pm$0.11& 1.7$\pm$0.5& 1.1$\pm$0.4\\
NGC\thin 4261   & 5.0 & 1.2$\pm$0.1& 1.0$\pm$0.1& 1.9$\pm$0.1& 1.3$\pm$0.1\\
NGC\thin 4472   & 4.3 & 0.51$^{+0.24}_{-0.31}$& 0.36$\pm$0.1& 1.3$\pm$0.3& 0.83$\pm$0.15\\
NGC\thin 4649   & 4.9 & 0.90$\pm$0.13& 0.65$\pm$0.12& 1.7$\pm$0.2& 1.1$\pm$0.1\\
NGC\thin 6482   & 2.9 & $0.73^{+0.18}_{-0.27}$& 0.52$^{+0.19}_{-0.23}$& 1.58$\pm$0.02& 1.05$\pm$0.02\\
\enddata
\tablecomments{K-band stellar mass-to-light ratios measured from our 
fits to the data using both the NFW+stars and the AC NFW+stars models. Since AC tends
to increase the cuspiness of the DM profiles, \mstars/\lk\ is substantially
lower for the AC NFW+stars models. We also show the predicted \mstars/\lk\ values
derived from simple stellar population synthesis, assuming either the
Salpeter  or \citet{kroupa01a} IMF.}
\end{deluxetable*}
%\clearpage
It is interesting to compare the stellar M/L ratios (\mstars/L) 
determined by 
our fitting to the expectations of stellar population synthesis
models. In order to ensure that the optical light traces the 
stellar mass as closely as possible, we opted to perform this
comparison in the K-band.
Table~\ref{table_mass_to_light}
shows \mstars/\lk\ determined from our models using eq~\ref{eqn_baryons}
to constrain \fbaryons.
Since AC tends to increase the cuspiness of the DM profile
we found a significantly lower mass-to-light ratio for the AC NFW+stars model
than for the NFW+stars model.

To compare our measured \mstars/\lk\ to single burst stellar population synthesis
predictions, we first estimated a mean emission-weighted
stellar age and metallicity for each galaxy, as 
outlined in Appendix~\ref{sect_stars}.
We linearly interpolated synthetic \mstars/\lk\ values based on
the stellar population models of \citet{maraston98a} from
updated model-grids made available by the author\footnote{\href{http://www-astro.physics.ox.ac.uk/~maraston/Claudia's_Stellar_Population_Models.html}{http://www-astro.physics.ox.ac.uk/$\sim$maraston/Claudia's}-\\\_Stellar\_Population\_Models.html}.
For typical early-type galaxies, K-band and 
\twomass\ \ks-band magnitudes should
differ by $<$0.1~magnitudes \citep{carpenter01a}, so we were
able to compare directly the synthetic K-band M/L ratios with
our measured \mstars/\lk\ ratios.
The predicted \mstars/\lk\
ratios are shown in Table~\ref{table_mass_to_light} for different
assumptions about the stellar IMF, which is poorly-known in 
early-type galaxies. In this case we show predicted \mstars/\lk\ assuming
a standard Salpeter IMF, and for the IMF of \citet{kroupa01a}.
It is immediately clear that these ratios
are very sensitive to this choice; \mstars/\lk\ is typically
$\sim$50--60\% higher if the Salpeter IMF is used.

Our measured \mstars/\lk\ for the NFW+stars models are typically
$\sim$20\% lower than the synthetic M/L ratios, assuming the 
Kroupa IMF. Using the AC NFW+stars models, the discrepancy is $\sim$40\%.
Assuming a Salpeter IMF, the discrepancies for both models
are considerably larger. This would seem to rule out the 
Salpeter IMF, in agreement with the conclusions of 
\citet{padmanabhan04a}. 
The best-fitting \mvir\ and c are sensitive to
\mstars/\lk. If we fix \mstars/\lk\ to the synthetic value, this 
essentially pushes all the galaxies, except NGC\thin 4261 (for which
the measured and synthetic values are in excellent agreement) and 
NGC\thin 720 in the direction of  the high-\mvir\ range of their 
confidence contours shown in Fig~\ref{fig_confidence}. 
For NGC\thin 720, \mvir\ is lowered and c increased. The 
fits are then typically much worse ($\Delta \chi^2 \sim$7--35),
and the loci in the \mvir-c plane slightly more discrepant with 
simulations.

There are a considerable number of systematic uncertainties
in the computation of the synthetic M/L ratios, not the least
of which is the very uncertain IMF, which could
probably account for the modest discrepancy with our NFW+stars
results (see \S~\ref{discussion_mass_to_light}).
In the case of NGC\thin 720, the rather young age inferred for
the stellar population ($\sim$3~Gyr) leads to a significantly lower
synthetic \mstars/\lk\ than measured. Fitting template models
to spatially-resolved spectra of this system, \citet{rembold05a}
found evidence of a significant age gradient, which falls from
$\sim$12~Gyr in the centre to $\sim$3~Gyr by 1~kpc. This may, therefore,
represent a system in which a relatively small fraction of the 
stellar component, produced in a modest,  recent star-formation
event (``frosting'') dominates the optical line emission. In this case,
the synthetic \mstars/\lk\ may be underestimated. We return to this issue
in \S~\ref{discussion_mass_to_light}.

\section{Systematic errors} \label{sect_systematics}
%\clearpage
\begin{deluxetable*}{lllllllllll}
\tablecaption{Systematic error budget}
\tabletypesize{\scriptsize}
\tablehead{ \colhead{Galaxy} & \colhead{Best-fit} & \colhead{$\Delta$stat} &
\colhead{$\Delta$N04} & \colhead{$\Delta$stars} & \colhead{$\Delta$bkd} & 
\colhead{$\Delta$asym} & \colhead{$\Delta$temp} & \colhead{$\Delta$spectra}
& \colhead{$\Delta$dist} & \colhead{$\Delta$baryons}
}
\startdata
\multicolumn{11}{c}{\mvir/$10^{12}$\msun}\\ \hline
NGC720 &$6.6$ &$^{+2.4}_{-3.0}$ &$+0.06$ &$^{+0.5}_{-0.3}$ &$-1.7$ &$-0.2$ &$-0.9$ &$^{+0.7}_{-0.2}$ &$^{+0.9}_{-0.6}$ &$-0.01$ \\
NGC1407 &$16$ &$\pm6$ &$-0.1$ &$^{+1}_{-0.5}$ &$-3$ &$+3$ &$^{+1}_{-3}$ &$^{+1}_{-3}$ &$^{+3}_{-4}$ &$-0.3$ \\
NGC4125 &$6.2$ &$^{+0.8}_{-2.3}$ &$-0.2$ &$^{+1.7}_{-2.1}$ &$^{+0.09}_{-0.2}$ &$^{+0.2}_{-0.03}$ &$-1.0$ &$^{+0.5}_{-2.0}$ &$^{+0.7}_{-0.6}$ &$-0.3$ \\
NGC4261 &$67$ &$^{+41}_{-15}$ &$+2$ &$^{+15}_{-24}$ &$+9$ &$^{+5}_{-0.5}$ &$^{+0.08}_{-0.6}$ &$^{+13}_{-0.04}$ &$^{+4}_{-7}$ &$-3$ \\
NGC4472 &$33$ &$^{+6}_{-10}$ &$+0.5$ &$^{+6}_{-0.05}$ &$+2$ &$+12$ &$-4$ &$^{+8}_{-1}$ &$\pm 3$ &$-5$ \\
NGC4649 &$35$ &$^{+7}_{-13}$ &$-8$ &$^{+16}_{-3}$ &$+2$ &$-1$ &$^{+63}_{-2}$ &$^{+6}_{-2}$ &$\pm 3$ &$-19$ \\
NGC6482 &$7.1$ &$^{+4.4}_{-1.7}$ &$-0.8$ &$^{+1.9}_{-0.3}$ &$-0.1$ &$+0.8$ &$^{+0.1}_{-3.6}$ &$^{+0.3}_{-0.8}$ &$^{+0.9}_{-0.8}$ &$-0.2$ \\
\hline\multicolumn{11}{c}{c}\\ \hline
NGC720 &$18$ &$^{+30}_{-8}$ &$+0.1$ &$\pm2$ &$+6$ &$+0.2$ &$+2$ &$^{+2}_{-1}$ &$\pm 2$ &$+0.03$ \\
NGC1407 &$18$ &$^{+11}_{-7}$ &$-0.8$ &$^{+2}_{-3}$ &$^{+4}_{-3}$ &$-4$ &$^{+4}_{-1}$ &$^{+6}_{-3}$ &$^{+5}_{-3}$ &$+0.02$ \\
NGC4125 &$10$ &$^{+5}_{-2}$ &$-1$ &$^{+5}_{-4}$ &$^{+0.9}_{-0.1}$ &$^{+0.05}_{-1.0}$ &$+2$ &$^{+2.}_{-0.7}$ &$^{+2}_{-1}$ &$+0.3$ \\
NGC4261 &$3.7$ &$\pm 1.7$ &$-1.4$ &$^{+3.6}_{-1.2}$ &$^{+0.10}_{-0.7}$ &$-0.3$ &$^{+0.08}_{-0.03}$ &$-1.5$ &$^{+0.8}_{-0.1}$ &$+0.08$ \\
NGC4472 &$13$ &$^{+4}_{-2}$ &$-2$ &$^{+0.4}_{-2}$ &$-0.7$ &$-5$ &$+1$ &$^{+0.2}_{-3.}$ &$\pm 2.$ &$+1.0$ \\
NGC4649 &$21$ &$^{+6}_{-3}$ &$-2$ &$^{+3}_{-6}$ &$-0.9$ &$+2$ &$^{+3}_{-4}$ &$^{+0.4}_{-3}$ &$\pm 3$ &$+7$ \\
NGC6482 &$18$ &$^{+13}_{-8}$ &$+1$ &$^{+2}_{-6}$ &$+3$ &$+6$ &$^{+51.}_{-0.4}$ &$+2$ &$\pm 3$ &$-0.7$ \\
\hline \multicolumn{11}{c}{\mstars/\lk (NFW+stars)}\\ \hline
NGC720 &$0.77$ &$^{+0.52}_{-0.71}$ &$+0.17$ &$^{+0.28}_{-0.18}$ &$-0.15$ &$+0.07$ &$-0.04$ &$\pm 0.10$ &$\pm 0.16$ &$-0.001$ \\
NGC1407 &$0.52$ &$^{+0.25}_{-0.32}$ &$+0.06$ &$^{+0.71}_{-0.16}$ &$^{+0.12}_{-0.04}$ &$+0.14$ &$-0.19$ &$^{+0.09}_{-0.18}$ &$^{+0.10}_{-0.08}$ &$+0.001$ \\
NGC4125 &$0.72$ &$\pm 0.11$ &$+0.04$ &$^{+0.62}_{-0.20}$ &$^{+0.01}_{-0.03}$ &$+0.06$ &$-0.04$ &$+0.05$ &$\pm 0.15$ &$-0.003$ \\
NGC4261 &$1.2$ &$\pm 0.1$ &$+0.05$ &$^{+0.7}_{-0.8}$ &$^{+0.008}_{-0.001}$ &$^{+0.04}_{-0.05}$ &$^{+0.0009}_{-0.006}$ &$^{+0.09}_{-0.010}$ &$\pm 0.2$ &$-0.006$ \\
NGC4472 &$0.51$ &$^{+0.24}_{-0.31}$ &$+0.06$ &$^{+0.45}_{-0.06}$ &$+0.05$ &$+0.20$ &$-0.01$ &$^{+0.13}_{-0.02}$ &$^{+0.11}_{-0.09}$ &$-0.01$ \\
NGC4649 &$0.91$ &$\pm 0.13$ &$+47$ &$^{+0.80}_{-0.19}$ &$+0.02$ &$-0.06$ &$-0.16$ &$^{+0.05}_{-0.009}$ &$\pm 0.18$ &$-0.03$ \\
NGC6482 &$0.73$ &$^{+0.18}_{-0.27}$ &$+0.13$ &$^{+0.50}_{-0.04}$ &$-0.06$ &$-0.15$ &$-0.51$ &$-0.05$ &$\pm 0.15$ &$-0.01$ \\
\hline \multicolumn{11}{c}{\mstars/\lk (AC NFW+stars)}\\ \hline
NGC720 &$0.54$ &$^{+0.42}_{-0.48}$ &$-0.12$ &$\pm 0.16$ &$-0.14$ &$+0.06$ &$-0.05$ &$^{+0.03}_{-0.10}$ &$^{+0.12}_{-0.09}$ &$-0.003$ \\
NGC1407 &$0.35$ &$\pm 0.25$ &$-0.06$ &$^{+0.41}_{-0.08}$ &$^{+0.12}_{-0.05}$ &$+0.12$ &$-0.14$ &$^{+0.07}_{-0.13}$ &$^{+0.07}_{-0.05}$ &$-0.001$ \\
NGC4125 &$0.53$ &$\pm 0.11$ &$-0.04$ &$^{+0.46}_{-0.14}$ &$^{+0.01}_{-0.03}$ &$+0.06$ &$-0.04$ &$+0.04$ &$^{+0.11}_{-0.09}$ &$-0.002$ \\
NGC4261 &$1.0$ &$\pm 0.1$ &$+0.02$ &$\pm0.6$ &$^{+0.02}_{-0.004}$ &$^{+0.03}_{-0.04}$ &$-0.006$ &$^{+0.1}_{-0.003}$ &$\pm 0.2$ &$-0.009$ \\
NGC4472 &$0.36$ &$\pm0.1$ &$-0.04$ &$^{+0.27}_{-0.03}$ &$+0.04$ &$+0.17$ &$-0.02$ &$^{+0.09}_{-0.01}$ &$^{+0.08}_{-0.06}$ &$-0.010$ \\
NGC4649 &$0.65$ &$\pm 0.12$ &$-0.05$ &$^{+0.57}_{-0.12}$ &$+0.02$ &$-0.06$ &$-0.13$ &$^{+0.06}_{-0.008}$ &$\pm 0.13$ &$-0.05$ \\
NGC6482 &$0.52$ &$^{+0.19}_{-0.23}$ &$-0.07$ &$^{+0.37}_{-0.03}$ &$-0.05$ &$-0.13$ &$^{+0.005}_{-0.03}$ &$-0.04$ &$^{+0.11}_{-0.09}$ &$-0.01$ \\
\enddata
\tablecomments{The estimated error-budget for each of the galaxies.
Excepting the statistical error, these values estimate a likely
upper bound on the sensitivity of the (best fit) value of each
parameter to various data-analysis choices, and should {\em not}
be added in quadrature with the statistical errors.
The systematic uncertainties on \mvir\ and c are estimated for
the NFW+stars model.
In addition to the best-fit values, we show the 90\% confidence
interval for each parameter ($\Delta$stat). We also show
estimated upper-limits on the systematics likely to arise by making
various changes to our default analysis choices. This includes
adopting the N04 DM model ($\Delta$N04), varying the shape of the 
stellar mass component ($\Delta$stars), varying the background
($\Delta$bkd), excluding data in the vicinity of asymmetries
($\Delta$asym), adopting alternative temperature models
($\Delta$temp), changing spectral analysis choices ($\Delta$spectra),
varying the distance ($\Delta$dist) or assuming that all of the
stellar baryons are in the central galaxy ($\Delta$baryons).} \label{table_syserr}
\end{deluxetable*}
%\clearpage
In this section we address the sensitivity of our results to various
data-analysis choices which were made. An estimated upper limit
on the sensitivity of our results to these choices is shown in
Table~\ref{table_syserr}. These numbers reflect the 
sensitivity in the best-fit parameter to 
each potential source of systematic error, and we stress they
should {\em not} be added in quadrature with the statistical 
errors. We outline in detail below how each of these 
systematics were estimated. Those readers
uninterested in the technical details of our analysis may like
to proceed directly to \S~\ref{sect_discussion}.

\subsection{DM profile shape} \label{sect_n04}
As discussed above, we experimented with replacing the NFW model by
the revised N04 model, which is less cuspy. 
We caution that the \mvir-c relation was 
derived assuming NFW.
We fixed the $\alpha$ parameter for this model to 0.17, the 
mean value determined from simulations since the inner slope 
of the DM halo is degenerate to some degree with the stellar mass. 
The quality of the fits (N04+stars, AC N04+stars) were typically similar to 
those using the simple NFW profile. There were some slight systematic 
differences
in the inferred \mvir\ and c as compared to NFW. It is interesting
to note that this model, which is less cuspy than NFW, gave
slightly larger \mstars/\lk, although not sufficiently to bring our
measurements completely into agreement with the synthetic estimates.
For the adiabatically-compressed N04 model, \mstars/\lk\ did not
increase, but this is unsurprising since the stellar component
significantly modifies the shape of the inner DM halo in this 
model.  We note that the predicted typical inner slope for
\lcdm\ halos is still under debate.  If, instead of the N04 profile
we had adopted the cuspier profile of  \citet{diemand05a}, 
then the resultant \mstars/\lk would have been even smaller, and in 
worse agreement with stellar population models.

\subsection{Shape of the stellar potential} \label{sect_syserr_potential}
To account for the stellar component, we adopted an H90 model,
the effective radius of the model being fixed to that 
determined by \twomass. However, it is not entirely clear
that the H90 model is an adequate description of the stellar
mass. There are some deviations between H90 and the 
de Vaucouleurs model fitted as the {\em de facto} standard
to the optical light profiles of elliptical galaxies,
particularly in the critical central regions. 
Furthermore, the K-band light profiles of elliptical galaxies may, 
in fact, be better described by the \sersic\ profile
\citep[\eg][]{brown03a}.
To investigate the sensitivity of our results to the H90 assumption,
therefore, we experimented with adopting a \sersic\ stellar mass
potential \citep[\eg][]{prugniel97a}. 
To determine the two parameters of this model 
(the \sersic\ index and half-light
radius) we obtained the \ks-band \twomass\ images of each galaxy
from \ned, and fitted the surface brightness profiles
using dedicated software. A \sersic\ model fitted the 
\ks-band light profile of each galaxy
in the radial range  5\arcsec--3\arcmin\ reasonably well.
The fitted profiles tended to be slightly more centrally peaked than 
H90, which resulted primarily in
slightly {\em lower} inferred \mstars/\lk\ ratios when adopted
as mass models. We also experimented with replacing the H90 model
with a de Vaucouleurs model \citep{mellier87a}, 
and adopting \reff\ values from \citet{pahre99a}.

Elliptical galaxies exhibit radial colour gradients, which may reflect
gradients in the metallicity or age of the stellar population
(see discussion in \S~\ref{discussion_mass_to_light}). These
may therefore imply a radial gradient in the stellar M/L ratio.
It is beyond the scope of this present work to take such a gradient
into account. However, we investigated the sensitivity of 
our results to the precise shape of the optical light profile
we adopted by experimenting with replacing the K-band \reff\
for each galaxy with the (typically larger) B-band value listed in RC3.
For NGC\thin 6482, for which \reff\ is not listed in RC3, we simply
increased \reff\ by 50\%.

\subsection{Background subtraction} \label{sect_systematics_bkd}
One of the major potential sources of systematic uncertainty
in measuring the mass profiles of galaxies is the background 
subtraction technique. This is especially important in the
low surface-brightness regime at large radii in our galaxies. 
In order to 
estimate the likely magnitude of uncertainty arising from
our modelling, when initially fitting the background
components (\S~\ref{sect_bkd}) we artificially adjusted the 
slope of the instrumental background components, which dominate
at high energy, to the limits
of their 90\% confidence regions, refitting the other components
and then refitted all the spectra with these revised background
models.

\subsection{X-ray asymmetries} \label{sect_asymmetry}
We note that there are some low-level asymmetries in the X-ray 
surface brightness profiles (\S~\ref{sect_imaging}). In order to
assess the potential impact of these features, we experimented
with excluding or including the features. In particular, we 
tried
excluding data from the vicinity of the jet and  AGN in 
NGC\thin 4261. We also excluded data from an off-axis X-ray 
asymmetry in NGC\thin 4125 and excluded data for NGC\thin 4472 
outside 6\arcmin, where \citet{irwin96} pointed out that the 
X-ray data become asymmetric.
These choices did not dramatically affect our results, indicating
that these features did not indicate a significant
violation of hydrostatic equilibrium, at least in an 
azimuthally-averaged sense. 
To gain an insight into possible asymmetries in other sources, we 
tried re-extracting all our spectra, and re-deriving the mass
profiles, from suitably-oriented semi-annuli
(thereby excluding one half of the emission from each system).

\subsection{Temperature profile}
In principle multiple temperature profiles may be able to fit the 
same data adequately well but give rise to slightly different
global halo parameters. In particular our constraints upon 
\fbaryons, the computation of which requires the  extrapolation of the density 
(and hence temperature) profiles to large radius, may make 
\mvir\ and c sensitive to this effect. To test this, we 
experimented with cycling through each of our adopted temperature profiles
(eq~\ref{eqn_trise2}--\ref{eqn_twophase2}). Provided the fits
were of comparable quality to our preferred choice, the impact on
the best-fit parameters reflect the systematic uncertainty in this 
choice. Furthermore, we also experimented with excluding the central
bin from the temperature profiles of NGC\thin 1407 and NGC\thin 4649,
which may indicate a central disturbance (although there is no
obvious X-ray morphological disturbance in this region).
These choices did not strongly affect our results.

\subsection{Spectral-fitting choices} \label{sect_systematics_spectra}
A variety of choices are made in the spectral-fitting, each of 
which can affect, to some degree, the inferred \rhog\ and temperature
in each radial bin. A thorough discussion of these 
effects is given in \citet{humphrey05a}.

{\em Column density.} In order to take account of possible
local deviations in the line-of-sight \nh\ from the value
of \citet{dickey90}, we experimented with allowing \nh\ to vary
by $\pm$25\%.

{\em Bandwidth.} To estimate the impact of the bandwidth on
our fits, we experimented with fitting the data in the energy
ranges 0.7--7.0~keV, 0.5--2.0~keV and 0.4--7.0~keV, in addition
to our preferred choice 0.5--7.0~keV.

{\em Plasma code.} There are some uncertainties in the correct
modelling of the individual emission lines, in particular those of Fe. This
can systematically lead to differences in the inferred temperature
and density depending on choice of plasma code. We therefore 
experimented with replacing the APEC model with the MEKAL plasma model.

{\em Unresolved source component.} We included a 7.3~keV bremsstrahlung
component to account for unresolved point sources 
within \dtwentyfive. This model is generally a good fit to the
resolved point sources in early-type galaxies, but is an 
empirical result which may not be appropriate to model
all unresolved sources in all early-type galaxies. We therefore 
tested the sensitivity of our results 
to this model, by replacing the bremsstrahlung component
with a simple power law (with $\Gamma=$1.5) or varying the temperature
of the component by $\pm$25\% %DONE

\subsection{Distance uncertainty}
The estimated distance to the object enters into our mass determination
(Eq~\ref{eqn_hydrostatic_rho}) primarily through the impact on the 
radial scale.
To assess its impact on our fitting,
we experimented by varying the distance by $\pm$20\%.

\subsection{Stellar baryon fraction} \label{sect_systematics_baryon_fraction}
In our analysis, we restricted \fbaryons\ to enable interesting 
constraints on \mvir\ and c. For the stellar contribution to the 
baryon fraction, we assumed that mass is divided among group members
following the B-band light, which is not formally correct since
\mstars/L ratios are very sensitive to the age of the stellar 
population. To estimate how much impact this makes to our fits,
we experimented with assuming that all the stellar mass is in the central
galaxy, which should place an upper limit on the uncertainty
arising from this choice.

\section{Discussion} \label{sect_discussion}
\subsection{Hydrostatic equilibrium}
Our fit results provide strong evidence that the gas is in
hydrostatic equilibrium in these systems. Despite highly
nontrivial temperature and density profiles, we were able to recover
smooth mass profiles in remarkably good agreement with
expectation for these systems, using two complementary
techniques. If the gas is significantly out of 
hydrostatic equilibrium, this would represent a remarkable 
``conspiracy'' between the density and temperature profiles. 
It is unsurprising that the gas is close to hydrostatic equilibrium
in these systems, since we took care to choose objects with relaxed
X-ray morphology. Based on N-body/ hydrodynamical analysis,
X-ray measurements are expected to give reliable constraints 
on the DM in systems without obvious substructure
\citep{buote95a}.

Further support for hydrostatic equilibrium is provided by 
the general agreement between our measured
\mstars/\lk\ ratios and those predicted by SSP models, 
coupled with the agreement between the measured \mvir-c relation 
and that expected.
Similarly a comparison between our results and masses determined 
from stellar dynamics provides even more evidence that our 
measured mass profiles are reliable. 
Dynamically-determined \mgrav/\lb\ within the B-band  \reff\ are 
typically found to be $\sim$4--10 \citep{gerhard01a,trujillo04a}.
We found \mgrav/\lb\ within the B-band \reff\
(taken from RC3 or \citealt{faber89}) for our systems ranged
from $\sim$3 to $\sim$8, in good agreement
with this result. 
For NGC\thin 4649, outside \reff\ there is excellent agreement
between our measured \mgrav/L profile and that obtained from globular 
cluster kinematics, although at small radii the X-ray data
lie $\sim$30\% lower (K.\ Gebhardt et al, in preparation).
\citet{vandermarel91a} constructed stellar kinematical models
for 5 galaxies in our sample (NGC\thin 720, NGC\thin 1407 and 
NGC\thin 4261, NGC\thin 4472 and NGC\thin 4649), under the assumption
of a constant M/L profile. Strictly speaking a direct comparison cannot
be made between their \mgrav/\lb\ measurements and our results since our
data indicate this assumption is incorrect. However, if we simply assume 
that these M/L ratios represent those integrated out to \reff, the 
X-ray inferred
masses vary from $\sim$40\% lower to $\sim$10\% higher than those
from kinematics. \citet{kronawitter00a}
report \mgrav/\lb$\sim$8$\pm1.5$ for NGC\thin 4472 within $\sim$50\arcsec,
at which radius our X-ray determined value is $\sim$50\% lower. The
discrepancies between the X-ray and dynamical masses are only modest
(the X-ray mass being on average $\sim$20\% lower), indicating that
the data must be close to hydrostatic equilibrium.  
Turbulence is expected to contribute only $\sim$10\% pressure support
in clusters, which are believed to be more turbulent than galaxies,
\citep{rasia06a}.
Therefore, on a case-by-case
basis, the observed differences are most likely a manifestation of the 
mass-anisotropy degeneracy \citep[\eg][]{dekel05a}.

\subsection{Mass profiles}
We obtained detailed mass profiles for 3 galaxies and 4 group-scale
systems, out to $\sim$10\reff. The data clearly show M/L profiles
which are $\sim$flat within \reff\ and rise considerably outside
this range. This confirms the presence of substantial DM in
at least some early-type galaxies and indicates that a stellar mass 
component dominates within $\sim$\reff. 
This is consistent with studies of 
stellar kinematics and similar to the mass decomposition analysis
of \citet{brighenti97a}.

The data are well-fitted by a model comprising a stellar
mass (H90) component and an NFW DM profile. 
Omitting the stellar 
mass component led to systematically poorer fits, smaller
\mvir\ and vastly larger c ($\gg$20), which are inconsistent with
the predictions of \lcdm. This effect is easy to understand--- if
we add a compact stellar mass component to an (extended) NFW profile,
we increase the mass in the core which, by definition, makes the
halo more concentrated.
However, it is not entirely clear whether this
effect, pointed out by \citet{mamon05a},
can completely account for the significantly steeper
\mvir-c relation found by \citet{sato00a}.
Based on our analysis of group-scale halos \citep{gastaldello06a}
we found that the inclusion of the stellar mass component 
does not have a strong effect on c in most systems with 
\mvir \gtsim 2$\times 10^{13}$\msun,
provided the data are fitted to
a sufficiently large fraction of \rvir.
The data did not allow us to distinguish statistically between
the simple NFW+stars model and 
scenarios in which the DM halo experiences adiabatic compression
due to star formation (however, see \S~\ref{discussion_mass_to_light}),
or the NFW profile was replaced with the alternative N04 profile.

Comparing our inferred \mvir\ and c  to the predictions of \lcdm\
we find general agreement. There is some evidence, however, that
the concentrations are systematically higher than one would 
expect, although the error-bars are typically large. 
Such a trend is also seen in our analysis of groups
\citep{gastaldello06a}. Whilst the slope  of the \mvir-c relation
therefore implied by our data is difficult to explain by varying the 
cosmological parameters within reasonable limits \citep{buote06b},
we suspect that the discrepancy can be resolved by taking into
account the selection function of our galaxies. 

Our
data were not selected in a statistically complete manner and, by choosing
objects with relatively relaxed X-ray morphologies we are probably
selecting objects which have not had a recent major merger.
This systematically biases us towards early-forming, hence higher
concentration halos. 
In fact, it is striking that all three {\em de facto} galaxies
in our sample are relatively isolated systems (\S~\ref{discussion_groups}).
Such systems preferentially might be expected to occupy high-c 
halos \citep{zentner05a},
which does appear to be the case for 2 out of 3 of the galaxies.
We will return to these issues in detail in \citet{buote06b}.

\subsection{Galaxies, Groups and Fossil Groups} \label{discussion_groups}
All three of the lowest-mass systems in our sample are very isolated
optically. NGC\thin 6482 matches the isolation criteria adopted
to identify so-called ``fossil groups'' \citep{khosroshahi04a}. 
NGC\thin 4125 and NGC\thin 720 are both listed as ``groups'' in 
G93, but closer inspection actually reveals they are also
very isolated. Excepting the central galaxy, only one of the putative
members of the NGC\thin 720 ``group'' listed in the G93
catalogue  \citep[which omits the dwarf galaxy population 
studied by][]{dressler86a},
actually lies within the projected \rvir\ (but outside 0.75$\times$\rvir)
and it is 2.4 magnitudes fainter in B than the central galaxy.
\citeauthor{dressler86a} remarked upon the optical isolation of this galaxy.
Of the two putative companion galaxies to NGC\thin 4125 given 
in G93 which lie within the projected \rvir\ (but outside 
0.67$\times$\rvir), both are much fainter (by 2.3 and 3.9 magnitudes,
respectively) in B than the central galaxy.
In contrast, the four remaining systems in our sample are much less
optically isolated. \citet{schindler99a} show the clear over-density of 
early-type galaxies around NGC\thin 4649 and NGC\thin 4472, and almost
60 group members are associated with these systems by G93.
\citet{gould93a} identified at least 10 members of the NGC\thin 1407
group, from the dynamics of which he inferred a mass broadly consistent
with our measured \mvir\ (\S~\ref{sect_ngc1407}).
\citet{helsdon03a} report 57 galaxies associated with the NGC\thin 4261
group within $\sim$1~Mpc projected radius, which is consistent with 
our measured \rvir.

Rather than an isolated galaxy \citet{khosroshahi04a} identify NGC\thin 6482
as a ``fossil group''. Fossil groups are group-sized X-ray halos centred on 
essentially a 
single elliptical galaxy \citep{ponman94,vikhlinin99a,jones03}. The typical 
interpretation of these objects is groups in which all of the \lstar\ members 
have merged. Confusingly, using almost the same selection criteria,
 \citet{osullivan04b} classify the galaxy NGC\thin 4555 as an 
``isolated elliptical
galaxy'' and posit a very different formation scenario. 
This object appears to be more massive than NGC\thin 6482;
the authors found \mgrav $\sim 3\times 10^{12}$\msun\ within 60~kpc
which, assuming an NFW profile with c$=$15 would imply 
\mvir $\sim 2\times 10^{13}$\msun. 
Nonetheless, both of these systems have more in 
common (both optically and in the X-ray) 
with each other, and the other isolated ellipticals in our sample,
than, for example, the  massive (\mvir\gtsim$10^{14}$\msun), hotter (kT$\sim$2~keV)
fossil groups considered by \citet{vikhlinin99a}. 
We suspect that the distinction made between ``isolated elliptical'' and 
``fossil group'' for these two systems is largely semantic,
and consider NGC\thin 6482 more properly an isolated  galaxy, too.

The clear division in the galaxy content of our sample clearly lends itself
to the nomenclature  ``galaxies'' for the three lowest-mass systems,
and ``groups'' for the others. Strikingly, this separation 
between galaxies and groups also appears consistent with a difference in 
temperature profiles  (\S~\ref{discussion_temp}).
That this distinction appears commensurate with
\mvir$\sim 10^{13}$\msun\ is suggestive that this mass-scale may be a useful
yard-stick with which to compare to other systems. 
The error-bars on our mass estimates are sufficiently large that
the 90\% confidence regions of several of the objects (notably
NGC\thin 720, NGC\thin 6482 and NGC\thin 1407) actually straddle 
$10^{13}$\msun. However, it is clear that {\em on average}, the systems
with \mvir \ltsim $10^{13}$\msun\ are galaxies. We note that the \mvir\ adopted here
is that {\em before}
any tidal truncation which is almost certainly occurring as NGC\thin 4472
and NGC\thin 4649 merge with Virgo (their untruncated \rvir\ would stretch
much of the distance to M\thin 87).
\mvir\ does not exactly correlate with formation epoch,
so that lower-mass halos may still be in the process of forming 
(hence contain multiple galaxies of similar magnitude), and  more 
massive halos may contain single, dominant ellipticals (fossil groups).
Nonetheless, classifying halos primarily on the basis of 
\mvir\ provides a straightforward way to locate them in the formation hierarchy.
Traditionally, galaxy-like and group-like systems are distinguished
on the basis of local over-densities of galaxies.  However,
placing optically-identified groups into a cosmological context
requires a firm understanding not only of the formation of DM halos
but also how galaxies populate them, which is much less
well-understood \citep[\eg][]{kravtsov04a}.
This problem is compounded by the difficulties faced by optical group-finding
algorithms in identifying very poor groups \citep[\eg][]{gerke05a}. Not only
can a significant fraction of putative groups be chance superpositions of 
galaxies, particularly along filaments, but adjacent groups can be merged,
such as happened for NGC\thin 4649 and NGC\thin 4472 in G93. If there are
only a few identified members, small-number statistics and the treatment of 
interlopers can affect their interpretation \citep[\eg][]{gould93a}.
To complicate matters further, some authors refer to 
{\em any} over-density of galaxies as a group, even a Milky Way-sized 
galaxy and its dwarf satellites \citep[\eg][]{tully05a}.

\subsection{Stellar Mass-to-Light Ratios} \label{discussion_mass_to_light}
Comparing our measured stellar M/L ratios
to the predictions of simple stellar population (SSP)
models, we found reasonable agreement provided one assumes
a \citet{kroupa01a} IMF. There is modest disagreement,
even when the less-cuspy N04 DM model was adopted.
Considering the uncertainties in the SSP modelling 
(discussed below), however, we believe this discrepancy 
is not significant.
If we allowed the DM 
profile to be modified by adiabatic compression, we obtained
substantially smaller \mstars/\lk\ values from our data,
(since it increases the cuspiness of the halo)
which are more discrepant with the SSP models. This result
casts doubt on AC being as significant an effect as currently modelled. 
However, the data
alone did not allow us statistically to distinguish between
the NFW+stars and AC NFW+stars models.
Nonetheless, this result is joining a growing body of literature
which similarly calls into question whether AC operates as
predicted \citep{zappacosta06a,kassin06a,sand04a}.

There are a number of major uncertainties in the 
computation of the stellar mass-to-light ratios from the SSP models.
Specifically, the results are very dependent upon the 
assumed IMF, which is not confidently known in early-type
galaxies.
Furthermore there is some evidence that early-type galaxies
frequently contain multiple stellar populations of different
ages, including a significant young population
\citep[\eg][]{rembold05a,nolan06a}. Depending on the mass
fraction of the young component, this may substantially 
reduce \mstars/\lk\ in the galaxy, hence possibly reconciling
the data and the AC NFW+stars model. 
A small amount of star formation may also give rise to a 
population of stars which can dominate the light in the galaxy
core, giving rise to significantly lower synthetic \mstars/\lk\
than measured. This may be the case in NGC\thin 720 (see 
\S~\ref{sect_mass_to_light}).
More problematically, there
are known to be significant abundance, or possible age,
gradients in the stellar populations of early-type galaxies 
\citep[\eg][]{trager00a,kobayashi99a,rembold05a}, which would translate
into stellar \mstars/\lk\ gradients. Our simple modelling
did not allow us to account for such an effect {\em per se}.
Although we suspect that such gradients
will primarily lead to a \mstars/\lk\ value which reflects
an average for the galaxy, \mstars/\lk\ does depend to 
some extent upon the shape of the assumed stellar potential.
Properly taking account of this effect is beyond the scope
of this present work, but may bring the synthetic M/L ratios
and our results into better agreement. Clearly this is only
one of a number of other systematic effects which may also
reconcile the slight discrepancy
(Table~\ref{table_syserr}).

\subsection{Baryon fractions}
An interesting result from our analysis is that these systems,
despite having masses \gtsim 5$\times 10^{12}$\msun,
do not appear in general to be baryonically closed.
To some extent this trend was enforced by applying 
Eq~\ref{eqn_baryons} to constrain the data. However, 
the excellent fits we obtained by this method, in
conjunction with 
the good agreement between the measured \mvir-c relation
and the predictions of \lcdm\ and, crucially,
our measurements at the group scale (which do not 
employ this restriction: \citealt{gastaldello06a}),
indicate that the inferred \fbaryons ($\sim$0.04--0.09; Table~\ref{table_results})
are accurate.
Furthermore, if we relaxed this constraint
and instead restricted \fbaryons\ to a finite range, we 
also found that the data tended to favour modest values of \fbaryons. In particular,
for any given system, the measured \mvir\ and \fbaryons\ were 
strongly anti-correlated, so that our upper \mvir\ constraint
is in part imposed by the {\em lower} limit we place on \fbaryons. 
Given the shapes of the \mvir-c contours (Fig~\ref{fig_confidence}),
it is clear that good agreement with the \mvir-c relation predicted
from simulations tends, therefore, to require rather modest values of \fbaryons.
This would suggest that strong feedback plays an important 
role in the evolution of these objects.

\subsection{Temperature profiles} \label{discussion_temp}
By inspection of the temperature profiles (Fig~\ref{fig_temp}) it
is immediately clear that, for all of the galaxy-scale systems
in our sample the temperature profiles have negative gradients.
In contrast the group-scale objects have positive temperature gradients,
similar to observations of other
X-ray bright groups and clusters
\citep{gastaldello06a,vikhlinin05b,piffaretti05a}. This radical difference 
in the temperature 
profiles seems consistent with our division of galaxies and groups 
at \mvir $\sim 10^{13}$\msun.
The origin of this distinct demarcation between objects around
$10^{13}$\msun\ is unclear, however.
Negative temperature gradients are expected for
isolated galaxies containing relatively cool gas, such as that
arising from stellar mass-loss.
In the deep stellar potential well,
compressive heating of the gradually inflowing gas can dominate 
over radiative cooling to produce a negative temperature slope.
In contrast, if hotter ($\sim$1--2~keV) baryons are allowed
to flow in, radiative cooling dominates to produce a positive 
temperature gradient \citep{mathews03a}. 
It is by no means clear, however, why the hot baryons appear
to be present only in the systems with \mvir\gtsim $10^{13}$\msun.
One possibility is the local environment;
all of the galaxy-scale objects are rather isolated, whereas
the groups NGC\thin 4472 and NGC\thin 4649, in particular,
are found in a relatively dense cluster environment, which could
provide a reservoir of hot baryons. However, such an explanation
cannot easily account for the positive temperature gradient in
NGC\thin 1407, which is comparatively isolated, or the 
isolated system NGC\thin 4555, which appears 
only slightly more massive than our galaxies.

It is possible that selection effects may have played some role in
the bimodal temperature profile behaviour, since both 
NGC\thin 4125 and NGC\thin 6482 are
classified in \ned\ as LINERS, and NGC\thin 720 has a dominant
young stellar population (Appendix~\ref{sect_stars}). 
However, none of these systems
show strong X-ray morphology disturbances in the core, 
which might indicate a substantial energy input from 
star-formation or AGN activity.
In any case, the cooling time in the core of NGC\thin 720 is only 
$\sim$200~Myr, substantially less than the 
implied time since the last major burst of star formation,
and so the negative temperature gradient cannot simply be
related to energy injection during a starburst. 
Furthermore, at least two of the group-scale
systems also harbour AGN and do not show obvious negative temperature
gradients in the core. 

Another example of an object we believe to be a galaxy
(rather than a group) which exhibits a negative temperature
gradient  is the S0 NGC\thin 1332 \citep{humphrey04b}.
A possible counter-example to this trend might by the 
``isolated elliptical galaxy'' NGC\thin 4555, which exhibits
a temperature profile akin to the groups in our sample 
\citep{osullivan04b}. However, as we discuss in 
\S~\ref{discussion_groups}, this probably has comparable \mvir\
to the groups.

Another intriguing feature of two of the group scale
objects is a central temperature peak, similar to a feature
we found in the cluster A\thin 644 \citep{buote05a}.
In that system, we found a significant offset between the 
X-ray centroid and the emission peak in an otherwise fairly
relaxed object. We suggested that both of these features
may be related to the cD ``sloshing'' in the potential well
of the cluster, which is relaxing following disturbance by,
for example, a merger. 
We do not find obvious evidence of a similar offset in either 
NGC\thin 1407 or NGC\thin 4649. However, these groups may 
be in a comparably more relaxed (evolved) state than A\thin 644.
Alternatively, the central peaks may be related to past AGN activity
heating the gas in the core of the galaxies, from which the system
has had time to relax dynamically but not cool completely.

\subsection{Is NGC\thin 1407 a ``dark group''?} \label{sect_ngc1407}
Based on the group member dispersion velocity \citet{gould93a}
suggested that NGC\thin 1407 may lie in a massive (\gtsim a few
$\times 10^{13}$\msun) DM halo. Although such a conclusion was
strongly dependent on the association of the galaxy NGC\thin 1400,
which exhibits a large peculiar velocity, with the group, we can
now confirm the presence of a substantial DM halo around this system.
Both the temperature profile and our best-fit mass are similar to the 
bright X-ray group NGC\thin 5044 \citep{buote06a}, and yet 
it is almost 2 orders of magnitude fainter in \lx.
NGC\thin 5044 appears to be close to baryonic closure 
\citep{mathews05a}, and so has likely retained most of 
its large gaseous halo. On the other
hand NGC\thin 1407 is not baryonically closed (we estimate
\fbaryons$\simeq$0.06) and so the loss of much of its hot gas
envelope easily explains its lower \lx/\lb. Since the masses
of the two systems are not considerably different, this 
points to substantial variation in the evolutionary
history of these two groups. In particular, feedback may
have operated more efficiently in evacuating the gas from
NGC\thin 1407.

\citeauthor{gould93a}'s preferred mass estimate 
($\sim10^{14}$\msun) would imply a remarkably high M/L ratio
for the system (\mvir/\lb$\sim$900\msun/\lsun), making NGC\thin 1407 
a bona fide ``dark group''. The existence of such an object 
would provide a valuable insight into the process of star 
formation in DM halos, as it would imply star formation was 
somehow inhibited in that system. This mass estimate is, however,
considerably larger than our preferred value
$\sim 1.5 \times 10^{13}$\msun, which implies a more modest
M/L ratio (\mvir/\lb$\sim$140\msun/\lsun). 
To some extent, though, our constraint
on \fbaryons, which was necessary to obtain interesting \mvir\
constraints,  has probably enforced this behaviour.
Such a restriction
may not be valid in a system with an unusual star-formation history
and so we experimented with freeing \fbaryons.
To enable \mvir\ to be constrained,  we
restricted c to lie on the 
best-fit \mvir-c relation found by \citet{bullock01a}. 
The best-fitting mass,
\mvir$=(9.7^{+17.8}_{-6.2})\times 10^{13}$\msun, was in good agreement 
with \citet{gould93a}'s values, but implies a baryon fraction
only of $\sim$0.003. Since this fit was statistically 
indistinguishable from the preferred model, we cannot
determine which mass estimate is more likely.

\section{Summary}
Using \chandra\ we have obtained detailed mass profiles 
centred on 7 elliptical galaxies,
of which 3 were found to have {\em de facto} galaxy-scale halos,
with \mvir $< 10^{13}$\msun, and 4 had group-scale 
($10^{13}$\msun$<$\mvir$< 10^{14}$\msun) halos.
These represent the best available data for nearby objects 
with comparable \lx. In summary:
\begin{enumerate}
\item The M/L ratio profiles were $\sim$flat within \reff\ and rose sharply
outside this region, indicating substantial DM in all 7 systems.
\item The data were well-described by a two component model, comprising
an NFW potential for the DM and a H90 stellar mass model. 
We were not able statistically to distinguish
between this scenario and one  in which the DM profile was modified by ``adiabatic
compression'' due to baryonic infall. Similarly, we could not 
distinguish between the NFW and the revised N04 DM halo profiles.
\item The distribution of the galaxies in the \mvir-c plane was
in broad agreement with the predictions of \lcdm, although with a 
slight trend toward more concentrated halos, in good agreement
with our modelling of X-ray bright groups and poor clusters
\citep{gastaldello06a}. 
This probably represents a galaxy
selection bias to earlier-forming systems, 
and we will discuss how we might account for it 
in \citet{buote06b}. Allowing AC to modify the shape of the DM halo
did not appreciably affect the \mvir-c relation.
\item Omitting the stellar mass component resulted in 
systematically poorer fits, smaller \mvir\ and unphysically large c, 
confirming the  conclusions of \citet{mamon05a}. This may explain
very large values of c found by some previous X-ray observers
\citep[\eg][]{sato00a,khosroshahi04a}.
\item For the NFW+stars model,  \mstars/\lk\ was found to be
in approximate agreement with the predictions of simple stellar
population synthesis models, assuming a \citet{kroupa01a} IMF. The AC NFW+stars
models have significantly lower \mstars/\lk\ which seems to cast
doubt on the AC scenario, although this conclusion is sensitive to
the considerable uncertainties in the theoretical modelling.
\item Despite having \mvir \gtsim 5$\times 10^{12}$\msun, typically 
\fbaryons $\sim$0.04--0.09 for each galaxy,
implying that feedback has played an important role in the evolution
of these systems.
\item The temperature profiles of the galaxy-scale systems all
exhibited negative radial gradients, whereas the group-scale objects
exhibited positive gradients, similar to the ``Universal'' temperature
profiles being found in other X-ray bright groups and clusters.
This implies a strict line of demarcation between systems at \mvir $\sim 10^{13}$\msun.
\item In two of the groups, we found central temperature peaks, similar
to that found in the cluster A\thin 644 \citep{buote05a}, but 
no obvious central disturbances in X-ray morphology. This may
relate to past AGN activity, following which the heated gas in the 
core of the galaxy has relaxed but not
cooled.
\item We confirm the suggestion of \citet{gould93a} that the 
elliptical galaxy NGC\thin 1407 lies at the centre of a massive
DM halo, possibly making it a ``dark 
group'' with an unusually large M/L.
Our best-fitting \mvir\ is considerably lower than that 
of \citeauthor{gould93a}, implying M/L more consistent
with normal groups.
Nonetheless, if we relax
the assumptions of our modelling very large masses 
(\mvir$\sim 10^{14}$\msun) are allowed.
\end{enumerate}

\begin{acknowledgements}
We would like to thank Oleg Gnedin for making available his adiabatic compression code. We would also like to thank  Karl Gebhardt for communicating with us 
results from his paper in preparation.
We  thank Louisa Nolan for interesting
discussions on the stellar populations of galaxies. 
This research has made use of data obtained from the High Energy Astrophysics 
Science Archive Research Center (HEASARC), provided by NASA's Goddard Space 
Flight Center.
This research has also made use of the NASA/IPAC Extragalactic Database (\ned)
which is operated by the Jet Propulsion Laboratory, California Institute of
Technology, under contract with NASA. 
In addition, this work also made use of the HyperLEDA database
(\href{http://leda.univ-lyon1.fr}{http://leda.univ-lyon1.fr}).
Support for this work was provided by NASA under grant 
NNG04GE76G issued through the Office of Space Sciences Long-Term
Space Astrophysics Program.
\end{acknowledgements}

\appendix
\section{Stellar population parameters} \label{sect_stars}
%\clearpage
\begin{deluxetable*}{lllllll}
\tablecaption{Stellar population parameters\label{table_ssp}}
\tabletypesize{\footnotesize}
\tablehead{
\colhead{Galaxy} & \colhead{indices} & \colhead{ref.} & \colhead{${\rm [\alpha/Fe]}$} & \colhead{age} & \colhead{${\rm [Z/H]_0}$}
 & \colhead{$\rm <[Z/H]>$} \\
\colhead{} & \colhead{} & \colhead{} & \colhead{} & \colhead{(Gyr)} & \colhead{} & \colhead{}}
\startdata
NGC\thin720$^\dagger$   & H$\beta$, Mgb, Fe5270, Fe5335  & 2 & 0.37$\pm0.05$ & 2.9$^{+1.3}_{-0.3}$ & 0.65$\pm$0.13& 0.48$\pm$0.18 \\
NGC\thin 1407$^\dagger$ &  H$\beta$, Mgb, Fe5270, Fe5335 & 1 & 0.33$\pm 0.02$ & 12$\pm2$ & 0.35$\pm0.06$ & 0.08$\pm0.06$\ddag     \\
NGC\thin 4125           & H$\beta$, Mgb, Fe5270, Fe5335  & 3 & 0.33$\pm 0.16$ & 13$\pm8$ & 0.16$\pm$0.25 & -0.11$\pm$0.25\ddag    \\
NGC\thin 4261           & H$\beta$, Mgb, Fe5270, Fe5335  & 2 & 0.25$\pm 0.02$ & 15$\pm1$ & 0.30$\pm0.03$ & -0.03$\pm$0.10         \\
NGC\thin 4472$^\dagger$ &  H$\beta$, Mgb, Fe5270, Fe5335 & 2 & 0.25$\pm0.03$ & 9$\pm2$   & 0.36$\pm0.05$ & 0.17$\pm$0.12          \\
NGC\thin 4649           & H$\beta$, Mgb, Fe5335          & 2 & 0.25$\pm0.02$ & 13$\pm2$  & 0.41$\pm0.04$ & 0.23$\pm$0.15          \\
NGC\thin 6482           &  Mgb, Fe5270, Fe5335           & 3 & 0.30$\pm0.15$ & 12          & 0.28$\pm0.15$ & 0.06$\pm0.15$\ddag   \\
\enddata
\tablecomments{Stellar population parameters determined from Lick
index fitting. The indices used in fitting are listed (indices),
as is the reference (ref) whence they were taken. 
Those mean stellar abundances (${\rm <[Z/H]>}$) marked
\ddag\ were estimated from the central abundance (${\rm [Z/H]_0}$)
adopting the mean abundance gradient ${\rm [Z/H]_0}-{\rm <[Z/H]>}=0.27$\citep[see][]{humphrey05a}. 
Where no error-bar
is given, the parameter was frozen. 
Table references: 1--- \citet{beuing02a}, 2--- \citet{trager00a}, 3--- \citet{trager98a} 
Results for galaxies marked $^\dagger$ were taken from \citet{humphrey05a}
}
\end{deluxetable*}
%\clearpage
The mass-to-light ratio of a stellar population is dependent upon both the
age and the metal abundance ([Z/H]) of the stars. To estimate these 
quantities  we searched the literature
to obtain Lick/IDS indices for each galaxy, which we fitted with
the simple stellar population (SSP) models of \citet{thomas03a}, using the
technique outlined in \citet{humphrey05a}. Briefly, we constructed 
a model by linearly interpolating the SSP models as a function of 
stellar age, metallicity and $\alpha$-element to Fe ratio, which was
then fitted {\em via} a $\chi^2$ minimization technique to 
those indices shown in Table~\ref{table_ssp}. \citet{trager00a} provided
indices measured in two apertures, which enabled us to take account 
of any abundance gradients, as outlined in \citet{humphrey05a}.
Where only a central Lick index was available, we estimated the total
emission-weighted abundance by correcting the central metallicity by
-0.27 dex. We did not attempt to take account of possible age gradients. 
The results, including the 
Lick indices adopted and the reference whence the indices were obtained,
are shown in Table~\ref{table_ssp}. 
This method implicitly assumes that all the 
stars were created in a single burst of star formation, which may
be over-simplistic if there are, in fact, multiple bursts of star 
formation in early-type galaxies \citep[\eg][]{rembold05a,nolan06a}.
We note that we were not able to obtain acceptable solutions for 
NGC\thin 6482 if we used the H$\beta$ index, which is the most sensitive
age indicator. This galaxy is classified in \ned\ as a LINER
and is rather blue for an old stellar population 
(${\rm M_B-M_K=3.4\pm0.2}$, whereas a 12~Gyr, solar abundance population is 
expected to have ${\rm M_B-M_K\simeq3.9}$: \citealt{maraston98a}).
% This is closer to a 0.6~Gyr colour.
Both of these factors might suggest the presence of a significant young
population of stars \citep[although see][]{cidfernandes04a}. 
It is beyond the scope of this present work, however,
to attempt to take account of this effect.

\bibliographystyle{apj_hyper}
\bibliography{paper_bibliography.bib}

\end{document}